\documentclass[useAMS,usenatbib]{mn2e}
\usepackage{hyperref}
\hypersetup{
     colorlinks   = true,
     citecolor    = blue,
     linkcolor     = blue
}
\usepackage{caption}

\usepackage[T1]{fontenc}
\usepackage{ae,aecompl}
\usepackage{natbib}
\def\fesc{\ifmmode f_{\rm esc} \else $f_{\rm esc}$\fi}

\usepackage{graphicx}	
\usepackage{amsmath}	
\usepackage{amssymb}	

\def\apj{ApJ}
\def\apjs{ApJS}

\def\aap{A\&A}
\def\aaps{A\&AS}
\def\aj{AJ}
\def\mnras{MNRAS}

\def\pasp{PASP}
\def\nat{Nature}

\title[The radiation energy balance]{The efficiency of ionising photon
production and the radiation energy balance
in compact star-forming galaxies}

\author[Y. I. Izotov et al.]{
Y. I.\ Izotov$^{1,2}$,\thanks{E-mail: izotov@mao.kiev.ua}
N. G.\ Guseva$^{1,2}$, 
K. J.\ Fricke$^{2,3}$,
C.\ Henkel$^{2,4}$,
\& D.\ Schaerer$^{5,6}$ 
\\
$^{1}$Main Astronomical Observatory, Ukrainian National Academy of Sciences,
27 Zabolotnoho str., Kyiv 03141, Ukraine\\
$^{2}$Max-Planck-Institut f\"ur Radioastronomie, 
                     Auf dem H\"ugel 
                     69, 53121 Bonn, Germany\\
$^{3}$Institut f\"ur Astrophysik, 
                     G\"ottingen Universit\"at, Friedrich-Hund-Platz 1, 
                     37077 G\"ottingen, Germany\\
$^{4}$Astronomy Department, King Abdulaziz University, 
                     P.O. Box 80203, Jeddah 21589, Saudi Arabia\\
$^{5}$Observatoire de Gen\`eve, Universit\'e de Gen\`eve, 
51 Ch. des Maillettes, 1290, Versoix, Switzerland\\
$^{6}$IRAP/CNRS, 14, Av. E. Belin, 31400 Toulouse, France
}

\date{Accepted XXX. Received YYY; in original form ZZZ}

\pubyear{2015}

\begin{document}
\label{firstpage}
\pagerange{\pageref{firstpage}--\pageref{lastpage}}
\maketitle

\begin{abstract}
We derive apparent and absolute ultraviolet (UV) magnitudes, and luminosities in the
infrared (IR) range of a large sample of low-redshift (0 $<$ $z$ $<$ 1) compact 
star-forming galaxies (CSFGs) selected from the Data Release 12 of the Sloan Digital 
Sky Survey (SDSS). These data are used to constrain the extinction law in the
UV for our galaxies and to compare the absorbed radiation in the UV range with 
the emission in the IR range. We find that the modelled
far- and near-UV apparent magnitudes are in good agreement with 
the observed {\sl Galaxy Evolution Explorer} ({\sl GALEX}) magnitudes.
It is found that galaxies with low and high equivalent widths EW(H$\beta$) of 
the H$\beta$ emission line require different reddening laws with steeper slopes 
for galaxies with higher EW(H$\beta$). This implies an important role of the 
hard ionising radiation in shaping the dust grain size distribution.
The IR emission in the range of 8 -- 1000 $\mu$m
is determined using existing data obtained by various infrared space telescopes.
We find that the radiation energy absorbed in the UV range is nearly equal
to the energy emitted in the IR range leaving very little room for
hidden star formation in our galaxies. 
Using extinction-corrected H$\beta$ luminosities and modelled SEDs in the
UV range we derive efficiencies of ionising photon production $\xi$ for the
entire sample of CSFGs. It is found that $\xi$ in CSFGs with high EW(H$\beta$) 
are among the highest known for low- and 
high-redshift galaxies. If galaxies with similar properties existed at redshifts
$z$ = 5 -- 10, they could be considered as promising candidates for the 
reionisation of the Universe.
\end{abstract}




\section{Introduction}\label{sec1}

Recent studies of low-redshift compact star-forming galaxies (hereafter CSFGs) 
revealed their important role in understanding star formation (SF) processes in 
galaxies at various redshifts. \citet{I16c} have shown that the SF in 
these 
galaxies is characterised by strong short bursts. This conclusion is based on
the difference of the relations between stellar mass $M_\star$
and H$\beta$ luminosity $L$(H$\beta$) for CSFGs with high and low equivalent 
widths EW(H$\beta$) of the H$\beta$ emission line.
Assuming bursting SF in CSFGs and correcting the $L$(H$\beta$) 
for the burst age removes differences of $M_\star$-- $L$(H$\beta$) relations 
for CSFGs with different EW(H$\beta$) \citep[compare Figs. 9a and 
9b in ][]{I16c} while the assumption of continuous SF fails to
reproduce the data.
The validity of a bursting scenario is further supported by a comparison
of {\sl Galaxy Evolution Explorer} ({\sl GALEX}) FUV-to-H$\beta$ and 
NUV-to-H$\beta$ luminosity ratios
for CSFGs with high and low EW(H$\beta$)s \citep[compare Figs. \ref{fig6} and 
\ref{fig8} in ][]{I16c}. Subsamples of those low-$z$ CSFGs are
``Green Pea'' (GP) galaxies at redshifts $\sim$ 0.1 -- 0.3 selected by
their intense green colour on Sloan Digital Sky Survey (SDSS) composite
images \citep{Ca09} and luminous compact galaxies (LCGs) in a larger redshift 
range selected using not only photometric but spectroscopic SDSS
data as well \citep*{I11}.

One of the most important features of CSFGs is that they 
very closely resemble in many respects high-redshift SFGs. 
In particular, their stellar masses, luminosities, and chemical 
composition are similar \citep{I14a}. Low-$z$ CSFGs and high-$z$ SFGs 
follow the same luminosity-metallicity, stellar mass-metallicity and 
star-formation rate (SFR) -- stellar mass relations \citep{I15}. 
Furthermore, strong emission lines are present
in the SDSS optical spectra of CSFGs, and their emission in the UV and 
optical ranges is dominated by the radiation of numerous young massive stars 
formed during the above mentioned short bursts of star formation \citep{I16c}.

The importance of studying CSFGs is strengthened by the 
discovery of the escaping ionising radiation in the Lyman continuum (LyC) in 
some of these galaxies with high 
[O~{\sc iii}]$\lambda$5007/[O~{\sc ii}]$\lambda$3727 
emission-line ratios \citep{I16a,I16b}. This finding and high efficiency 
of ionising photon production \citep{S16} support the idea that 
their counterparts at redshifts $z$ $\sim$ 5 -- 10 were likely the main sources 
of the reionisation of the Universe.

\begin{figure}
\includegraphics[angle=-90,width=0.99\linewidth]{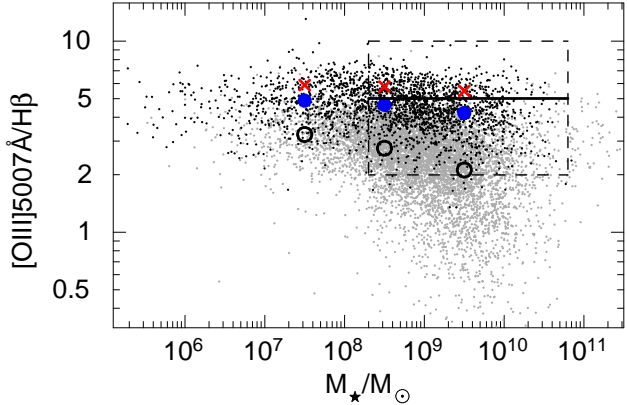}
\caption{The dependence of the [O~{\sc iii}]$\lambda$5007/H$\beta$
flux ratio on the stellar mass $M_\star$. Compact SFGs (CSFGs) with 
EW(H$\beta$)~$\geq$~50\AA\
and EW(H$\beta$)~$<$~50\AA\ are shown by black and grey dots, respectively.
The region delineated by a dashed line indicates ranges of 
[O~{\sc iii}]$\lambda$5007/H$\beta$
and $M_\star$ for a sample of Lyman-break galaxies (LBGs) at 
$z$ $\sim$ 3.5 with a median
value of [O~{\sc iii}]$\lambda$5007/H$\beta$ shown by a solid line
\citep{Sc13,T14,H16}.
Average values of [O~{\sc iii}]$\lambda$5007/H$\beta$
in 1~dex bins of $M_\star$ are shown by red crosses 
(EW(H$\beta$) $\geq$ 150\AA), blue filled circles
(EW(H$\beta$) $\geq$ 50\AA), and black open circles (EW(H$\beta$)~$<$~50\AA),
respectively.
\label{fig1}}
\end{figure}

A large sample of CSFGs opens an opportunity to study relations between
dust extinction in the UV range and its emission in the infrared (IR) range 
for the galaxies with various stellar masses, luminosities, and metallicities
by means of spectral energy distribution (SED) fitting of the SDSS spectra 
and their extrapolation to the UV. In particular,
\citet{I16a,I16b} showed that the extrapolation to the UV range of the SEDs 
obtained from the optical SDSS spectra
reproduces FUV and NUV {\sl HST} Cosmic Origin Spectrograph (COS) spectra 
reasonably well adopting the \citet*{C89} reddening law and
the ratio of the absolute-to-selective extinctions
$R(V)$ = $A(V)$/$E(B-V)$ $<$ 3.1. However, this result was obtained only
for five extreme CSFGs, which were observed with the {\sl HST}/COS,
and needs to be confirmed on larger samples of CSFGs.

The relations between interstellar extinction, UV and IR emission in low- and 
high-redshift SFGs were extensively discussed in the past 
\citep*[e.g. ][]{M99,C94,C00,R10,R12,B12,F16,Sh16}. However,
no such study was conducted for large samples of CSFGs.
One of the advantages of these galaxies is their compactness, which
ensures that aperture corrections needed to adjust spectroscopic and 
photometric observations are relatively small allowing to combine both sets
of data. Thus, different apertures do not introduce biases by comparing optical
spectroscopic data with UV and IR photometric data.

\begin{figure*}
\hbox{
\includegraphics[angle=-90,width=0.48\linewidth]{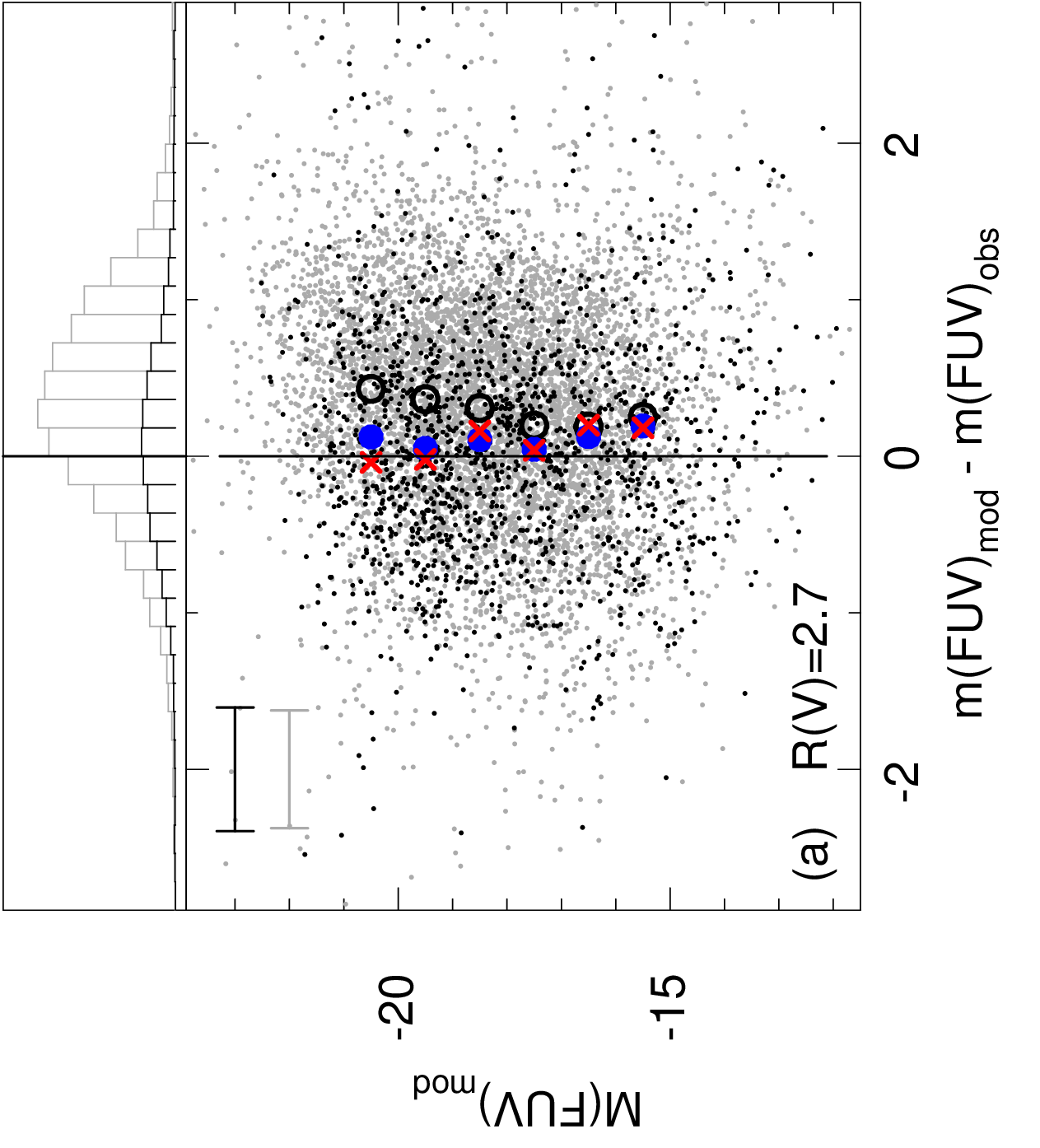}
\hspace{0.3cm}\includegraphics[angle=-90,width=0.48\linewidth]{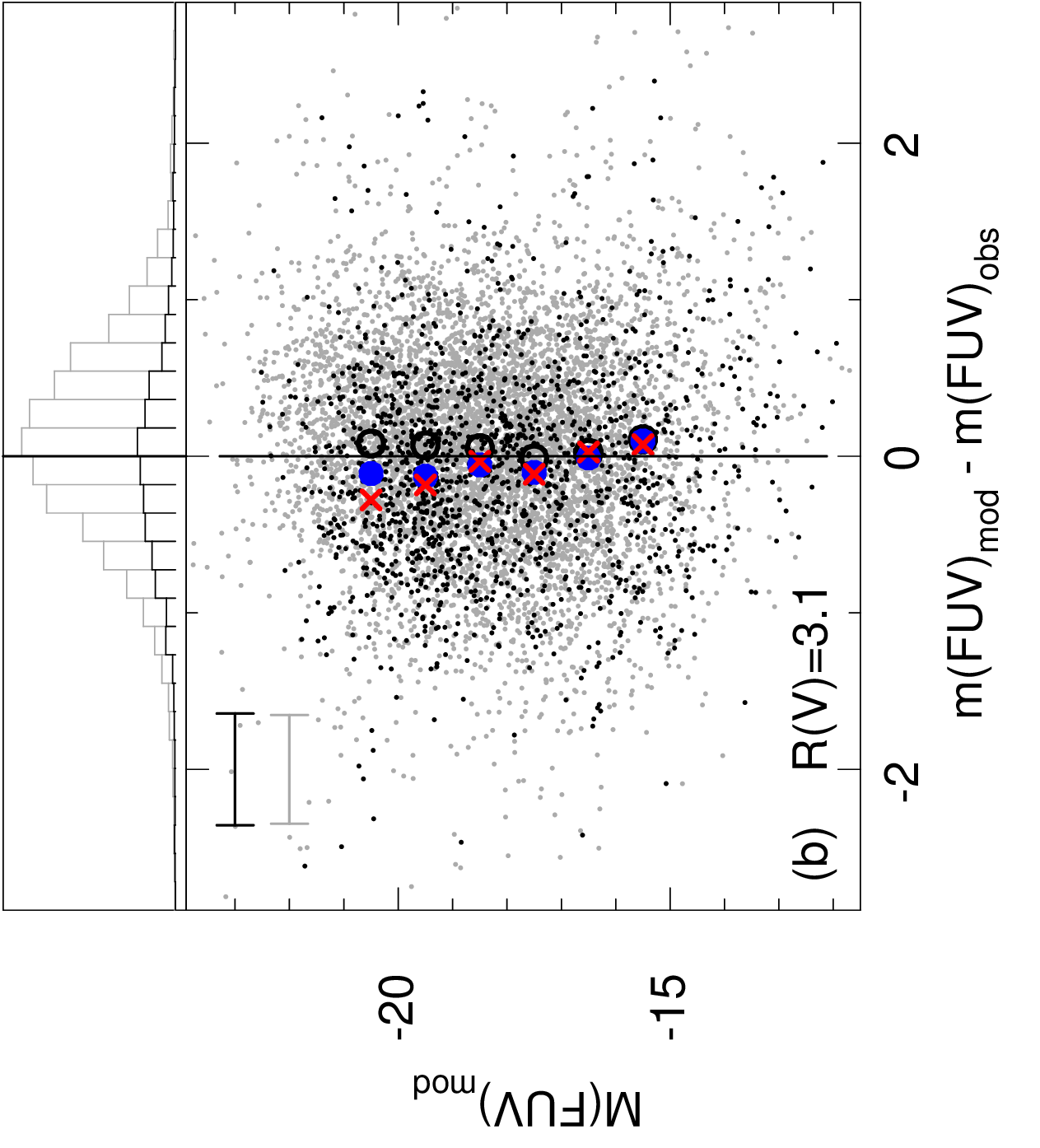}
}
\hbox{
\includegraphics[angle=-90,width=0.48\linewidth]{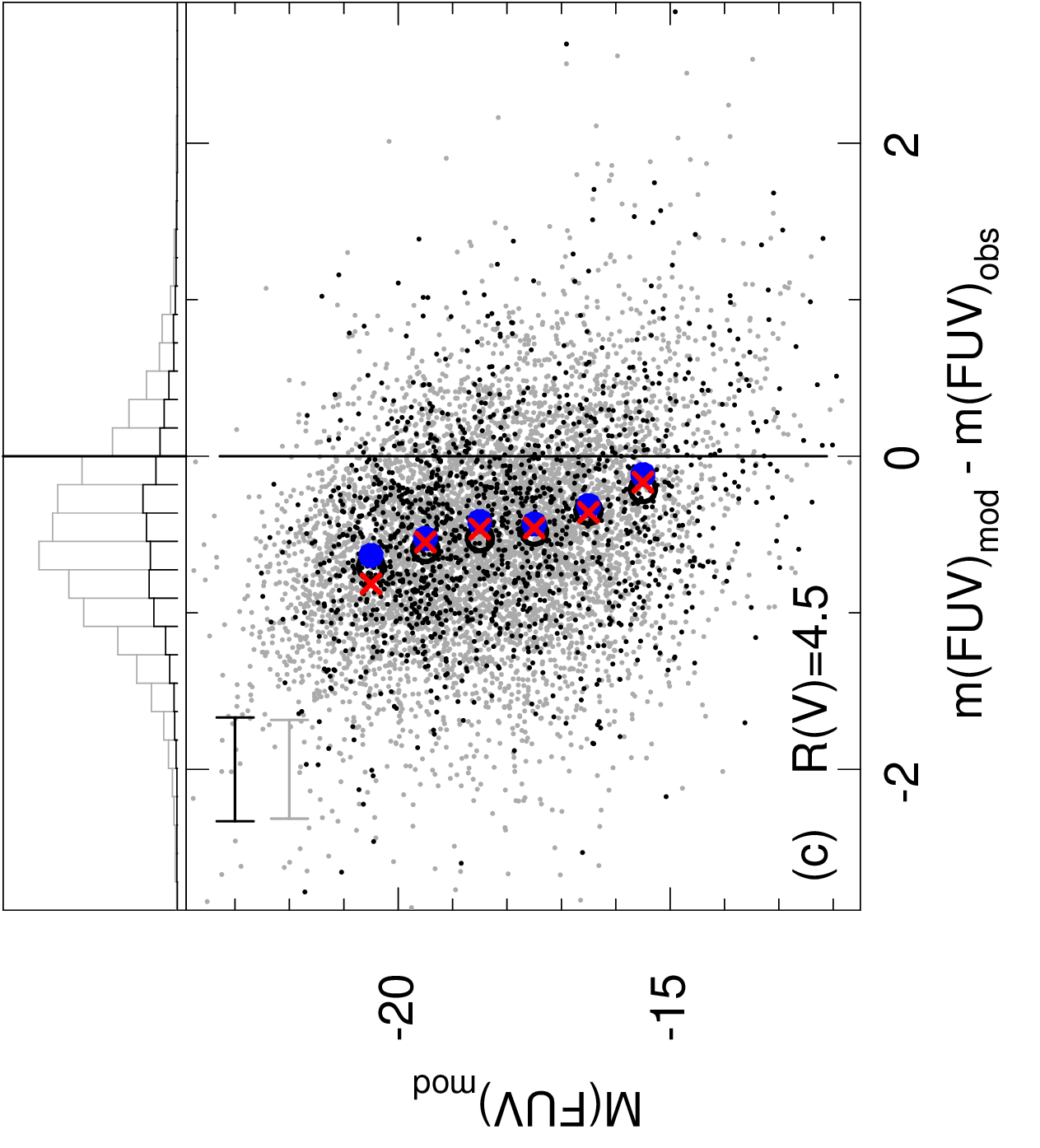}
\hspace{0.3cm}\includegraphics[angle=-90,width=0.48\linewidth]{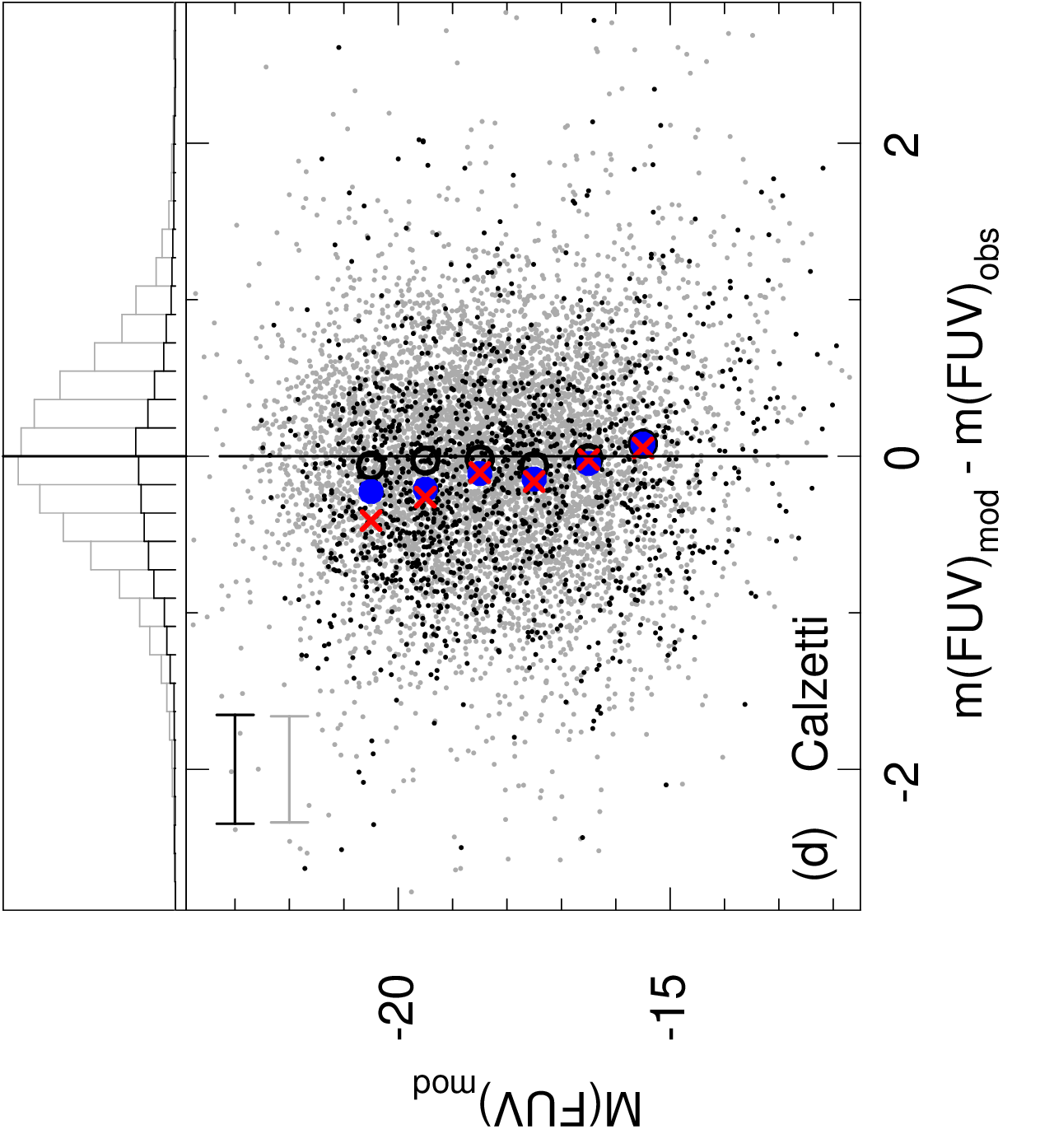}
}
\caption{The dependence of the difference between modelled
and {\sl GALEX} observed apparent FUV magnitudes 
$m$(FUV)$_{\rm mod}$ -- $m$(FUV)$_{\rm obs}$ on the modelled rest-frame
absolute FUV magnitude $M$(FUV)$_{\rm mod}$ (a-c) adopting the 
\citet{C89} reddening law with
various $R(V)$s and (d) the \citet{C94} reddening law. Galaxies with
EW(H$\beta$) $\geq$ 50\AA\ and $<$ 50\AA\ are shown by black and grey
dots, respectively. Average values of $m$(FUV)$_{\rm mod}$ -- $m$(FUV)$_{\rm obs}$ 
in 1 mag bins of $M$(FUV)$_{\rm mod}$ are shown by red crosses 
(EW(H$\beta$) $\geq$ 150\AA), blue filled circles
(EW(H$\beta$) $\geq$ 50\AA), and black open circles (EW(H$\beta$) $<$ 50\AA),
respectively.
On top are histograms of 
$m$(FUV)$_{\rm mod}$ -- $m$(FUV)$_{\rm obs}$ distributions for the entire
samples with EW(H$\beta$) $\geq$ 50\AA\ and $<$ 50\AA\ (black and grey lines,
respectively). Vertical lines in all panels indicate equal values of 
$m$(FUV)$_{\rm obs}$ and $m$(FUV)$_{\rm mod}$ and horizontal bars are 1$\sigma$ 
dispersions of $m$(FUV)$_{\rm mod}$ -- $m$(FUV)$_{\rm obs}$.
\label{fig2}}
\end{figure*}

The aim of this paper is threefold. First, we study the ability of
SED fitting of the SDSS spectra to reproduce the observed apparent
magnitudes of our sample galaxies in the UV range. We obtain slopes of 
the modelled SEDs in the UV range for further comparison with those of other
samples consisting of low- and high-$z$ SFGs and to study, which factors 
determining these slopes are most important, extinction or starburst age.
Second, using extinction coefficients and H$\beta$ emission-line fluxes obtained
from the SDSS spectra we derive the efficiency of ionising photon production of
CSFGs, an important parameter in studies of cosmic reionisation. 
Third, we wish to compare
luminosities of the absorbed UV radiation derived from the SED modelling with
the luminosities of SFGs in the IR range and to test whether any signs of 
hidden SF are present in CSFGs. We also wish to derive IR excesses, the ratios 
of IR and UV luminosities for the entire sample of CSFGs, and to compare them 
with those for other samples of low- and high-$z$ SFGs.

In Sect. \ref{sec2} we discuss the sample, modelling of the SEDs and
the determination of galaxy integrated parameters.
The comparison of the modelled and observed galaxy UV
magnitudes and the determination of the most appropriate reddening
law is discussed in Sect. \ref{sec3}. We derive
the intrinsic and observed SED slopes in the UV range and the efficiency
of ionising photon production in Sects. \ref{sec4} and \ref{sec5}. 
In Sect. \ref{sec6} the radiation absorbed in the UV is
compared with the radiation emitted in the IR. We derive IR excesses for CSFGs
and compare them with those for other samples of low- and high-$z$ SFGs.
The main results of the paper are summarised in Sect. \ref{sec7}.

\section{The sample of CSFGs}\label{sec2}


   A sample of CSFGs was selected from the SDSS Data Release 12 (DR12)
\citep{A15}. Selection criteria were described by \citet{I15}. Two main
criteria were compactness and stellar origin of ionising radiation. 
Only galaxies with
an angular radius on the SDSS images $R_{50}$ $\leq$ 3$''$
were selected, where $R_{50}$ is the galaxy's Petrosian radius within which 
50\% of the galaxy's flux in the SDSS $r$ band is contained.
Furthermore, all selected galaxies occupy the region of SFGs 
below the empirical \citet{K03} SFG -- AGN dividing line on the 
emission-line diagram 
[O {\sc iii}]$\lambda$5007/H$\beta$ -- [N {\sc ii}]$\lambda$6584/H$\alpha$
by \citet*{BPT81}. \citet{I16c} selected $\sim$ 14000 galaxies with redshifts 
$z$ $<$ 1 using the criteria mentioned above. The BPT diagram of CSFGs is
shown in Fig. 1 of \citet{I16c}. All these galaxies are characterised by
spectra with narrow emission lines and blue continua. The He {\sc ii} emission 
line in spectra of selected CSFGs is very weak or absent, indicating that
hard nonthermal ionising radiation is weak or absent in these galaxies.
We use this sample in the present study.

The SDSS photometric and spectroscopic data were supplemented by
{\sl GALEX} photometric data in the
UV range, {\sl Wide-field Infrared Survey Explorer} ({\sl WISE}), 
{\sl Infrared Astronomical Satellite} ({\sl IRAS}) and {\sl Herschel} 
photometric data in the IR range to derive SEDs, stellar masses, luminosities
and to compare observed and modelled apparent magnitudes in the far-UV (FUV)
and near-UV (NUV) ranges.

Monte Carlo calculations were used to derive intrinsic SEDs in the optical
range. The method was described e.g. by \citet{I11,I15}. SF history in CSFGs was
approximated by a recent instantaneous burst and by continuous SF with a 
constant SFR for old stellar populations. Both stellar and nebular 
continua and line emission were taken into account in the SED modelling. 
For fitting of the SEDs we use SDSS spectra corrected for
extinction with the reddening law by \citet{C89} with $R(V)$ = 3.1.
Varying $R(V)$ in the range 2.7 -- 4.5 results in differences less than a few 
percent between the SDSS spectra for our relatively transparent 
CSFGs \citep{I16c}. 
For stellar SEDs we adopted a Salpeter IMF \citep{S55}, Padova
evolutionary tracks \citep{G00} and a combination of stellar atmosphere models 
by \citet*{L97} and \citet*{S92}. The output quantities of the Monte
Carlo modelling are ages and masses of the young burst and the old stellar 
population with which SDSS spectra are best reproduced.

Aperture- and extinction-correction, and transformation of fluxes
to the luminosities were described in \citet{I15}.

The general properties of our SDSS DR12 sample of CSFGs are discussed
by \citet{I16c}. In many respects they are similar to the properties of
high-$z$ SFGs. In addition, we show in Fig. \ref{fig1} the distribution of
the [O~{\sc iii}]$\lambda$5007\AA/H$\beta$ emission-line ratio on the galaxy
stellar mass. Spectra of most CSFGs have high line ratios indicating active
ongoing SF and young starburst ages. 

Our galaxies are well
overlapping with $z$ $\sim$ 3.5 Lyman-break galaxies (LBGs) \citep{Sc13,T14,H16} implying similar
physical conditions in their ISMs. However, because of proximity the 
distribution of CSFGs extends to lower luminosities and stellar masses 
by almost three orders of magnitude. 
There is a slight increase of the [O {\sc iii}]/H$\beta$ and then its 
slight decrease with the stellar mass for galaxies with high EW(H$\beta$). This 
behaviour is predicted by photoionisation H {\sc ii} region models and is 
explained by higher metallicity and hence by the lower electron temperature in 
H {\sc ii} regions of more massive galaxies. The decrease of 
[O {\sc iii}]/H$\beta$ at high stellar masses is more prominent for galaxies 
with low EW(H$\beta$). It is caused not only by the lower electron temperature 
in H {\sc ii} regions of more massive galaxies, but also by considerable 
softening of ionising radiation at higher metallicities in bursts with high age 
(or low EW(H$\beta$)), which is insufficient to produce ion O$^{2+}$ in large 
quantities.

\section{Reddening in the UV range}\label{sec3}

\subsection{Modelled apparent magnitudes in the UV range}

We use extrapolations of SEDs derived from the optical SDSS spectra to
obtain apparent FUV magnitudes\footnote{http://asd.gsfc.nasa.gov/archive/galex/FAQ/counts\_back ground.html}:
\begin{equation}
m({\rm FUV})_{\rm mod}=-2.5\log\int{\frac{I(\lambda)Q_{\rm FUV}(\lambda) d\lambda}
{10^{[1+f({\lambda_0})]C({\rm H}\beta)}}} - 18.31,
\label{eq:mFUV}
\end{equation}
where $I(\lambda)$ is the flux of the intrinsic SED at the observed wavelength 
$\lambda$ attenuated by the (small) Milky Way extinction adopting a \citet{C89}
reddening law with $R(V)$ = 3.1 and expressed in 
erg s$^{-1}$ cm$^{-2}$ \AA$^{-1}$,
$C$(H$\beta$) is the galaxy's intrinsic extinction coefficient derived from the 
hydrogen Balmer decrement after correction for the Milky Way extinction,
which depends on $A(V)$ and $R(V)$ \citep[see Eq.~1 in ][]{I16a}, 
$\lambda_0$ = $\lambda$/(1+$z$) is the rest-frame wavelength,
$f(\lambda_0)$ = 0.4$\times$[$A(\lambda_0)$/$C$(H$\beta$) --1], and
$Q_{\rm FUV}(\lambda)$ 
is the transmission of the {\sl GALEX} FUV 
filter.\footnote{http://svo2.cab.inta-csic.es/svo/theory/fps3/index.php?id= GALEX/GALEX.FUV\&\&mode=browse\&gname=GALEX\& gname2=GALEX\#filter}

Similarly, the apparent NUV magnitudes are derived as
\begin{equation}
m({\rm NUV})_{\rm mod}=-2.5\log\int{\frac{I(\lambda)Q_{\rm NUV}(\lambda) d\lambda}
{10^{[1+f({\lambda_0})]C({\rm H}\beta)}}} - 19.17,
\label{eq:mNUV}
\end{equation}
where $Q_{\rm NUV}(\lambda)$ is the transmission of the {\sl GALEX} NUV 
filter.\footnote{http://svo2.cab.inta-csic.es/svo/theory/fps3/index.php?id= GALEX/GALEX.NUV\&\&mode=browse\&gname=GALEX\& gname2=GALEX\#filter}

\begin{figure}
\includegraphics[angle=-90,width=0.99\linewidth]{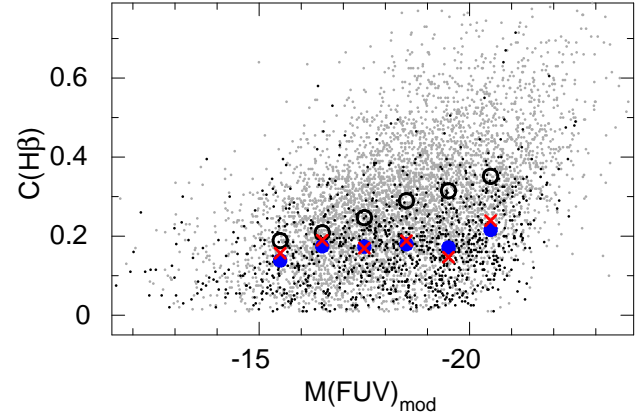}
\caption{The dependence of the extinction coefficient $C$(H$\beta$) on the 
modelled rest-frame absolute FUV magnitude $M$(FUV)$_{\rm mod}$. Symbols are
same as in Fig. \ref{fig2}.
\label{fig3}}
\end{figure}

\begin{figure*}
\hbox{
\includegraphics[angle=-90,width=0.48\linewidth]{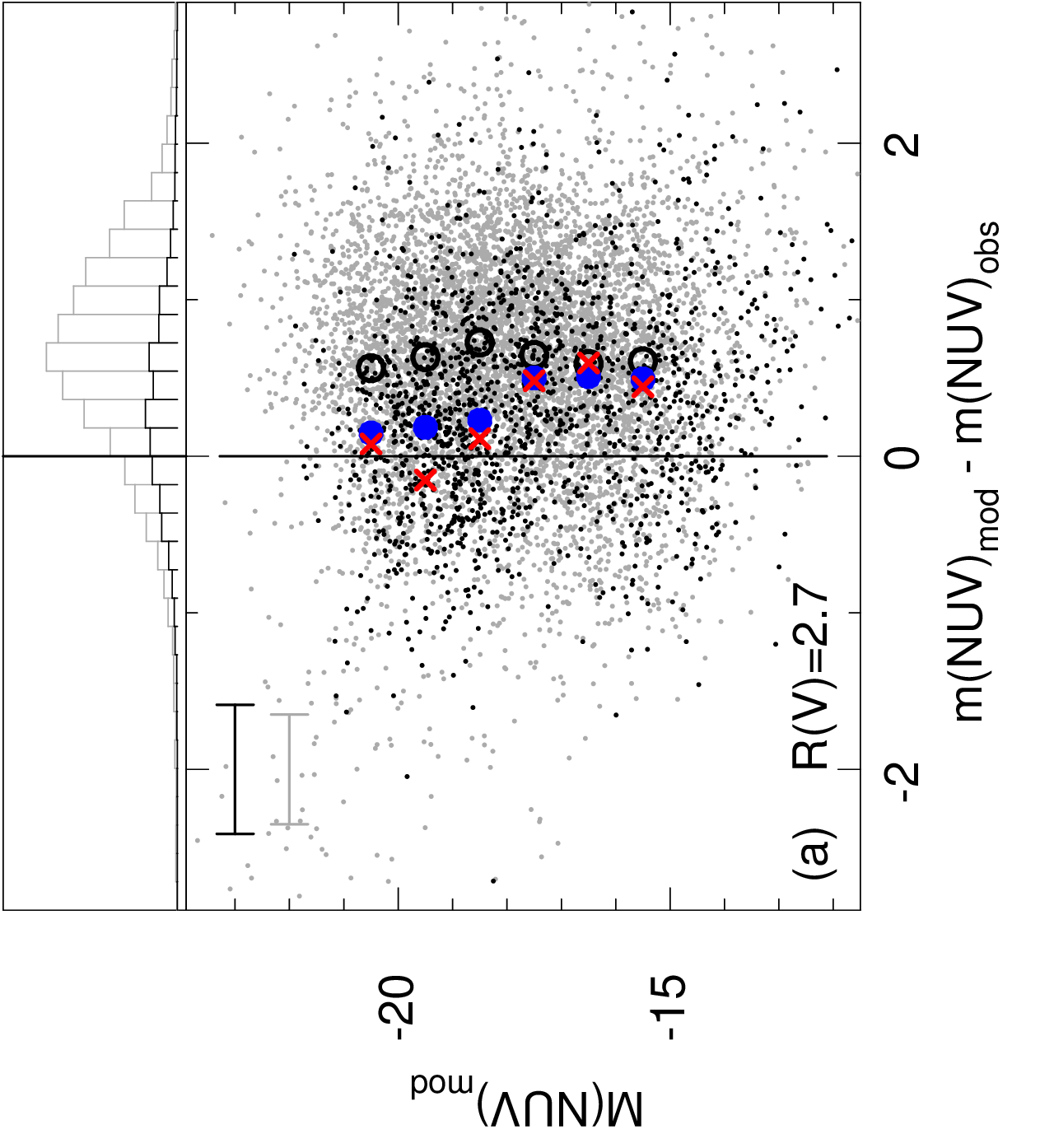}
\hspace{0.3cm}\includegraphics[angle=-90,width=0.48\linewidth]{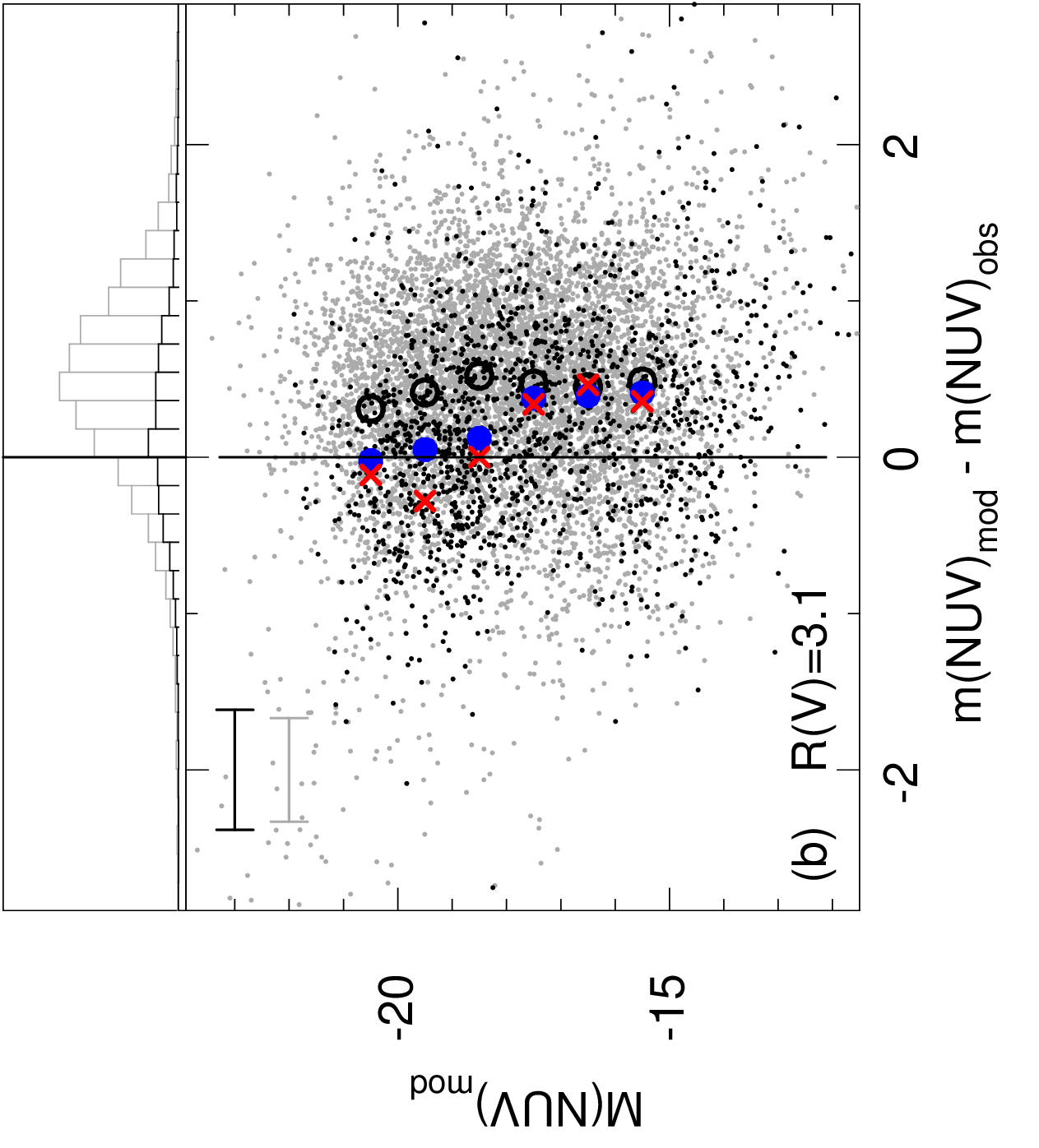}
}
\hbox{
\includegraphics[angle=-90,width=0.48\linewidth]{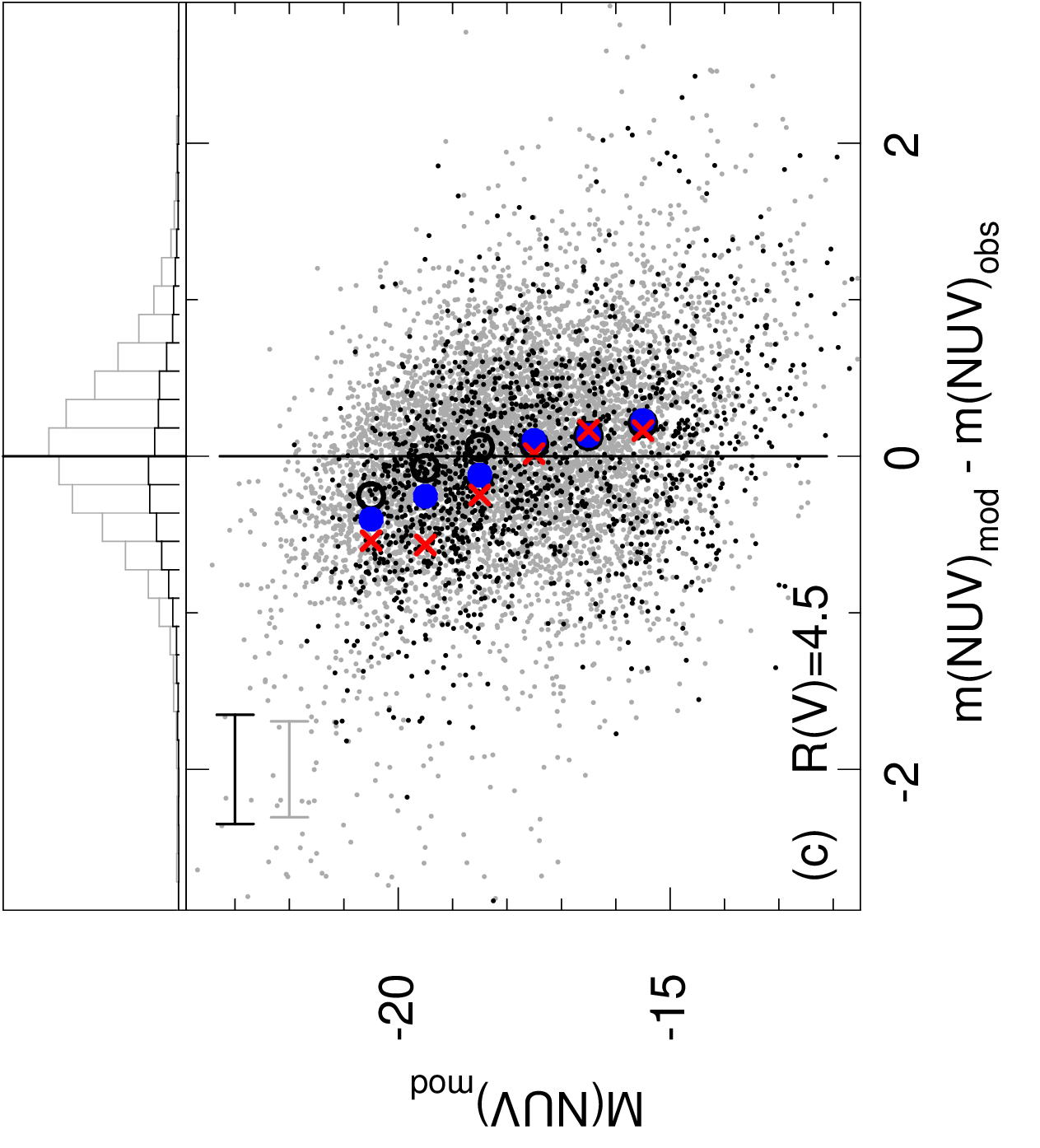}
\hspace{0.3cm}\includegraphics[angle=-90,width=0.48\linewidth]{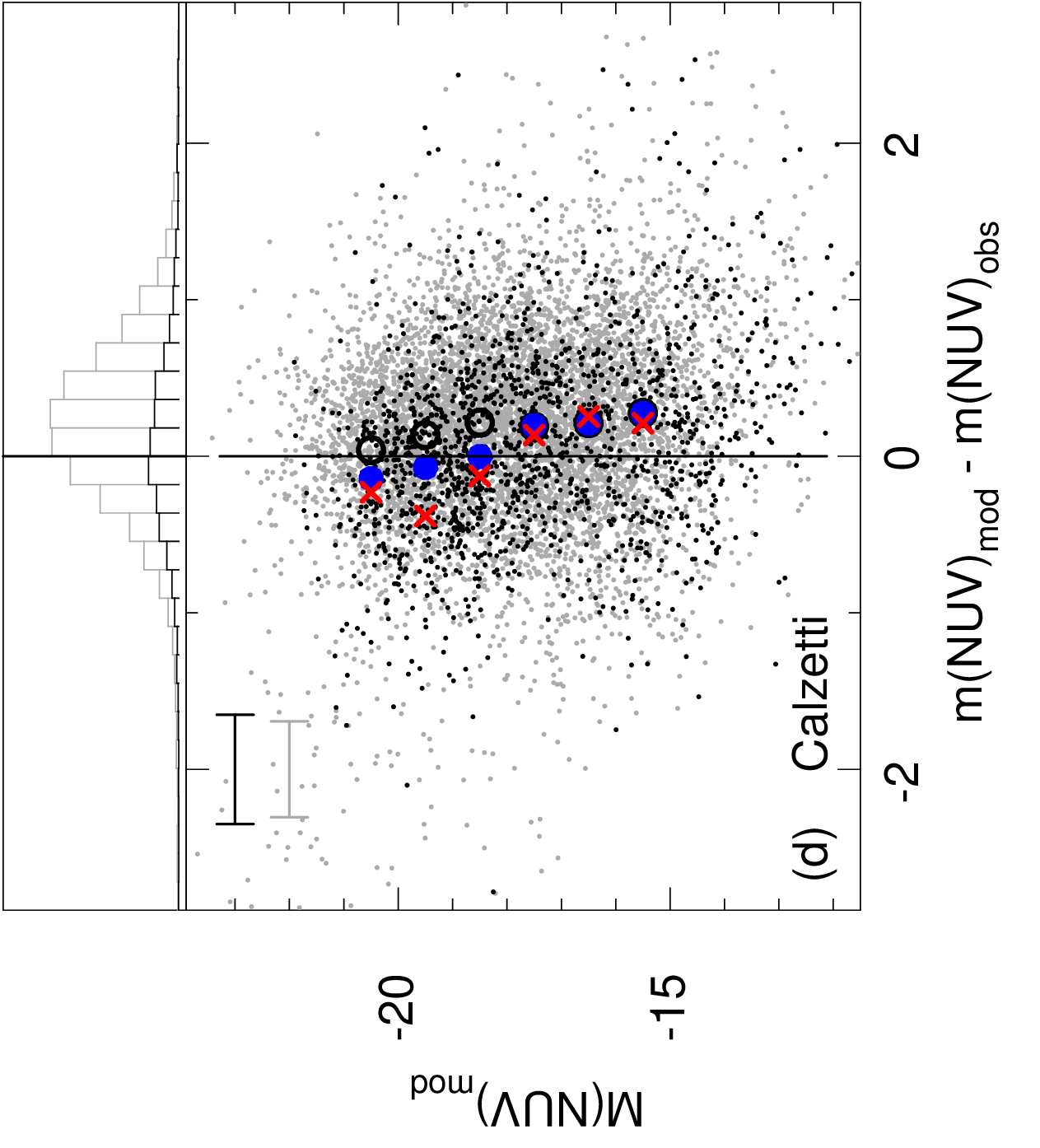}
}
\caption{Same as in Fig. \ref{fig2} but for NUV magnitudes. 
\label{fig4}}
\end{figure*}

\begin{figure}
\includegraphics[angle=-90,width=0.99\linewidth]{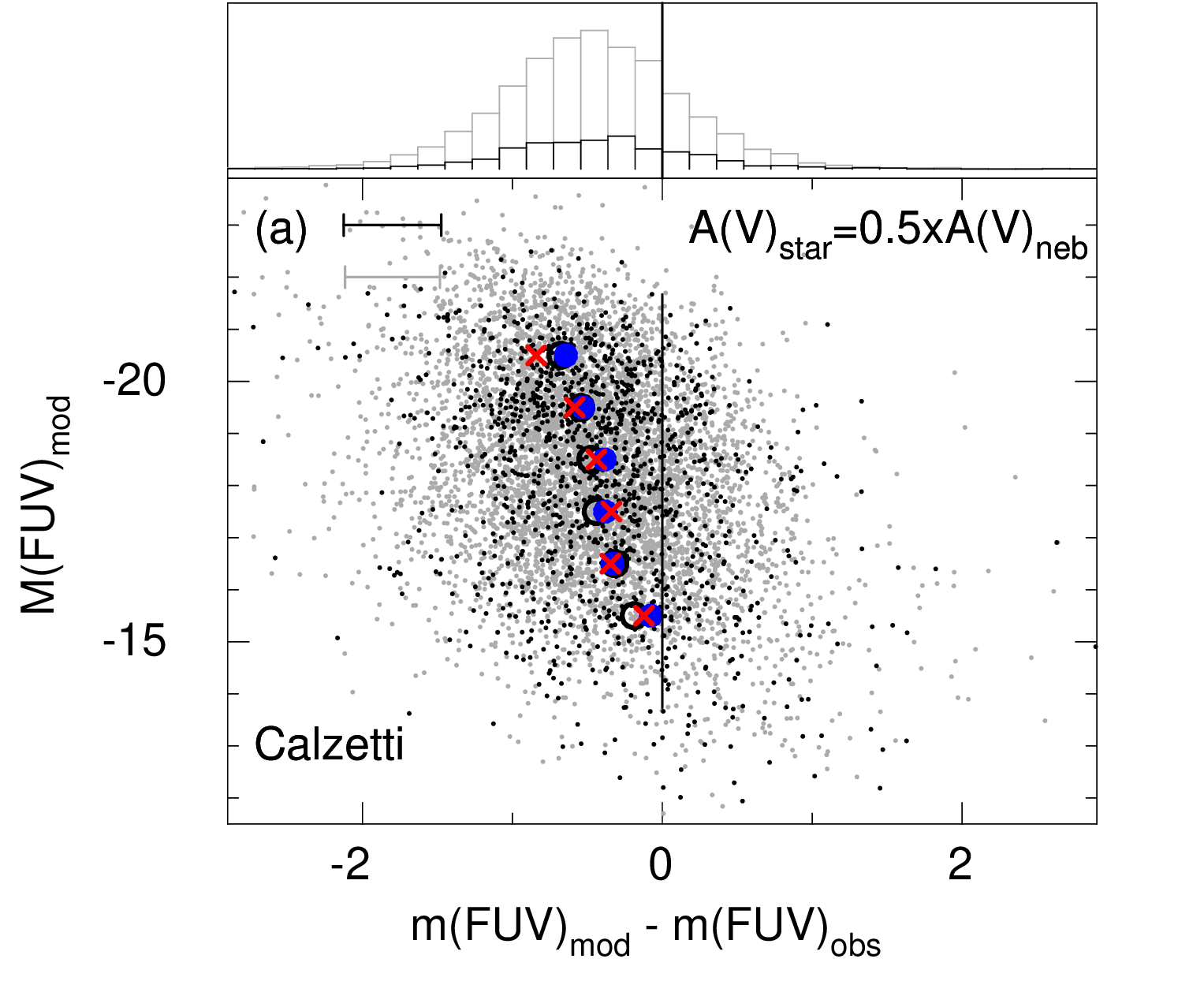}
\includegraphics[angle=-90,width=0.99\linewidth]{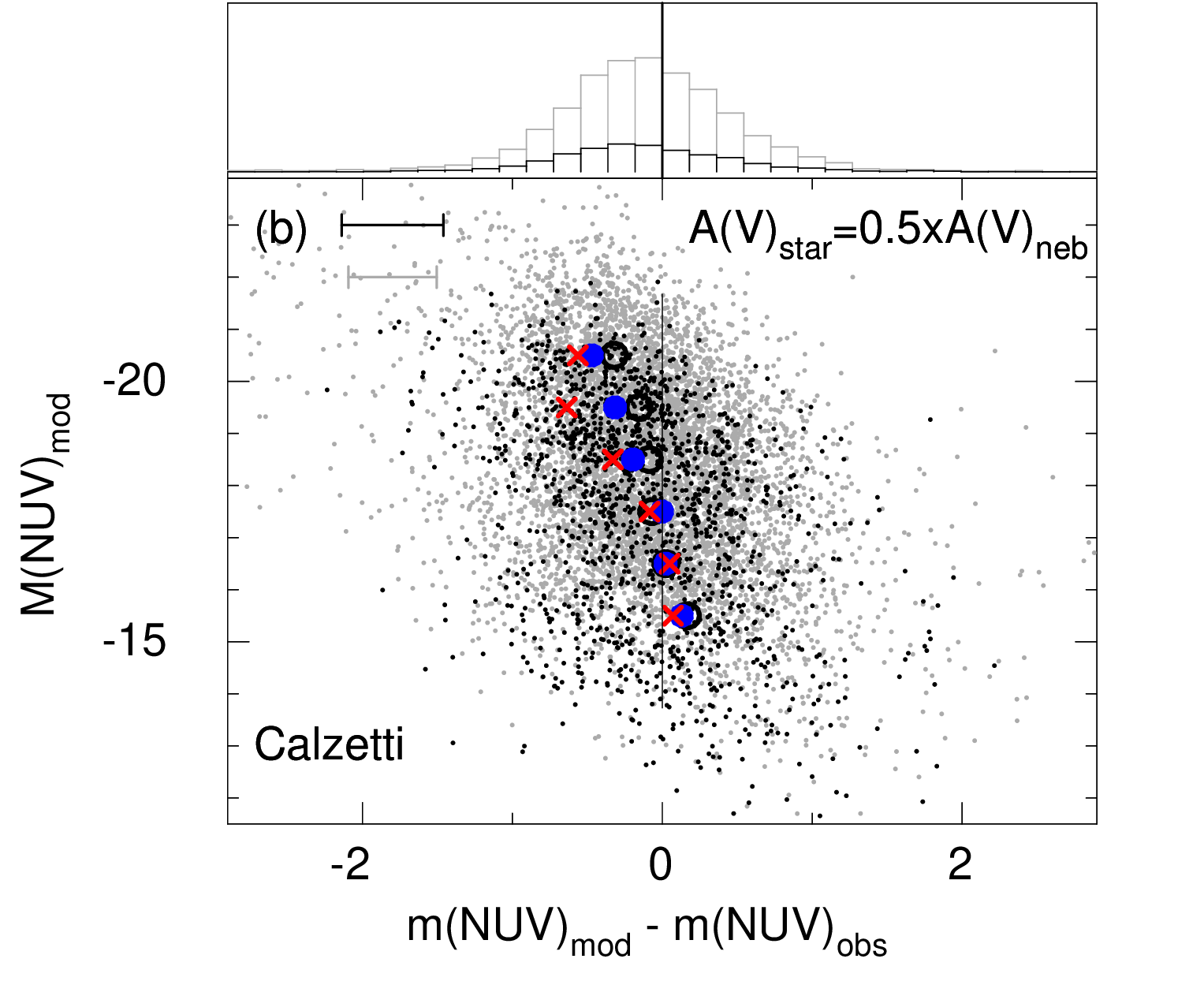}
\caption{(a) The dependence of the difference between modelled
and observed apparent FUV magnitudes 
$m$(FUV)$_{\rm mod}$ -- $m$(FUV)$_{\rm obs}$ on the modelled rest-frame 
absolute FUV magnitude $M$(FUV)$_{\rm mod}$ adopting the \citet{C94} reddening law
and stellar extinction equal to half of the nebular extinction. (b)~The same
for the modelled rest-frame absolute NUV magnitude $M$(NUV)$_{\rm mod}$.
For the symbols, see Fig. \ref{fig2}.
\label{fig5}}
\end{figure}

The rest-frame extinction-corrected absolute magnitudes are obtained as
\begin{eqnarray}
M({\rm FUV})_{\rm mod}=& -2.5 \log \int{I_0(\lambda_0) Q_{\rm FUV}(\lambda_0) d\lambda_0} \nonumber \\
                      & - 43.31 + 5\log D~~~~~~~~~~~~~~~~~
\label{eq:MFUV}
\end{eqnarray}
and
\begin{eqnarray}
M({\rm NUV})_{\rm mod}=& -2.5 \log \int{I_0(\lambda_0) Q_{\rm NUV}(\lambda_0) d\lambda_0} \nonumber \\
                      & - 44.17 + 5\log D,~~~~~~~~~~~~~~~~~
\label{eq:MNUV}
\end{eqnarray}
where $I_0(\lambda_0)$ is the flux of the intrinsic SED at the rest-frame 
wavelength $\lambda_0$, $D$ is the luminosity distance 
derived with a cosmological 
calculator \citep[NASA Extragalactic Database (NED),][]{W06}, based on the cosmological 
parameters $H_0$=67.1 km s$^{-1}$Mpc$^{-1}$, $\Omega_\Lambda$=0.682, 
$\Omega_m$=0.318 \citep{P14}.

\subsection{Comparison of modelled and observed apparent FUV and NUV magnitudes}

The aim of this Section is to study how well modelled FUV and NUV 
apparent magnitudes reproduce the observed data. We also wish to put constraints
on the reddening law in the UV range.

In Fig. \ref{fig2} we show dependencies of the differences between modelled
and observed apparent FUV magnitudes on the absolute FUV magnitudes
for $\sim$ 7400 CSFGs detected by {\sl GALEX} in the FUV range, 
where modelled magnitudes $m$(FUV)$_{\rm mod}$ are derived from 
the intrinsic SEDs, which are attenuated adopting extinction coefficients 
$C$(H$\beta$) 
and the \citet{C89} reddening law with various
$R(V)$s (Figs. \ref{fig2}a-c) and the \citet{C94} reddening law 
(Fig. \ref{fig2}d) (hereafter obscured SEDs). In all panels, the sample is 
split into galaxies with 
EW(H$\beta$) $\geq$ 50\AA\ (black dots) and with EW(H$\beta$) $<$ 50\AA\
(grey dots). The respective average values of 
$m$(FUV)$_{\rm mod}$ -- $m$(FUV)$_{\rm obs}$ 
in 1 mag bins of $M$(FUV)$_{\rm mod}$ are shown by blue filled and black open
circles. Additionally, by red crosses we show average values for the youngest
starbursts with EW(H$\beta$) $\geq$ 150\AA.
Equal modelled and observed apparent magnitudes are indicated by
vertical lines. On top of each panel are histograms of the
$m$(FUV)$_{\rm mod}$ -- $m$(FUV)$_{\rm obs}$ distribution for
CSFGs with EW(H$\beta$) $\geq$ 50\AA\ (black line) and EW(H$\beta$) $<$ 50\AA\ 
(grey line). 

It is seen in Fig. \ref{fig2} that $m$(FUV)$_{\rm mod}$ most
strongly depends on $R(V)$ for intrinsically brighter galaxies with 
EW(H$\beta$) $<$ 50\AA. This appearance can be 
explained by the fact that the average extinction in these galaxies is higher
than in fainter galaxies with EW(H$\beta$) $<$ 50\AA\ 
(black open circles in Fig. \ref{fig3}). This is consistent with conclusions 
made by e.g. \citet{P15} for GOODS-N galaxies at $z$~$\sim$~2.
We also note that average extinction in our CSFGs with higher
EW(H$\beta$) (blue filled circles and red crosses in Fig. \ref{fig3}) is lower 
and weakly depends on the galaxy intrinsic brightness.
 
The agreement between the observed and modelled magnitudes
for galaxies with high EW(H$\beta$) $\geq$ 50\AA\ (blue filled circles)
is essentially the same for
reddening laws with $R(V)$ = 2.7 and 3.1 excluding CSFGs
with $M$(FUV)$_{\rm mod}$ fainter than $-$16 mag (Figs. \ref{fig2}a,b).
However, since average $m$(FUV)$_{\rm mod}$ -- $m$(FUV)$_{\rm obs}$ values
for $R(V)$ = 2.7 and 3.1 are located to the right and to the left from the
line of equal magnitudes, respectively, 
the most appropriate value for $R(V)$ would be between 2.7 and 3.1.
Considering only galaxies with youngest starbursts as determined by their
high EW(H$\beta$) $\geq$ 150\AA\ we conclude that a reddening law with 
$R(V)$ = 2.7 is more preferable.
This is supported by SED fitting of the {\sl HST}/COS spectra in combination
with SDSS spectra of five LyC escaping galaxies with high 
EW(H$\beta$) $\sim$ 200\AA. \citet{I16a,I16b} showed that the observed
data are best fitted with the obscured SEDs adopting an extinction law
somewhat steeper in the FUV range than the widely used law with $R(V)$ = 3.1. 
On the other hand, the agreement is worse if the  
\citet{C94} reddening law is adopted (Fig.~\ref{fig2}d) with further worsening 
for a shallower reddening law with $R(V)$ = 4.5 (Fig.~\ref{fig2}c).

The agreement for galaxies with low EW(H$\beta$) < 50\AA\ (black open circles)
is the best if the \citet{C89} reddening law with $R(V)$ = 3.1 or the 
\citet{C94} reddening law are adopted (Figs. \ref{fig2}b,d).

Steepening of the reddening curve, corresponding to lower $R(V)$, 
with increasing EW(H$\beta$) may imply that
intense UV radiaton, which is higher in galaxies with high EW(H$\beta$)s,
modifies the size distribution of dust grains indicating 
a larger fraction of small dust grains. 
Higher EW(H$\beta$)s imply younger ages of starbursts and correlate with
higher ionisation parameters and thus with higher [O~{\sc iii}]/[O~{\sc ii}]
emission-line flux ratios. The modelled apparent
FUV magnitudes of galaxies with high [O {\sc iii}]/O {\sc ii}] better
reproduce observed magnitudes if low $R(V)$s are adopted. This is consistent
with the conclusions made for low-redshift Lyman continuum leaking galaxies 
\citep{I16a,I16b}. 

Somewhat different conclusions can be drawn from dependencies of differences 
between modelled and observed NUV magnitudes on absolute NUV magnitudes 
for $\sim$ 7900 CSFGs detected by {\sl GALEX} in the NUV range 
(Fig. \ref{fig4}). At variance with the FUV wavelength range, the reddening
curve in the NUV range in the case of \citet{C89} approximations 
is non-monotonic and includes the bump at 2175\AA.
It is seen that a better agreement between models and
observations for the CSFGs can be achieved with the \citet{C94} reddening law
or with the \citet{C89} reddening law with $R(V)$ = 4.5, 
excluding bright young starbursts shown by red crosses, where $R(V)$ = 2.7 is
more appropriate, similar to the FUV range.
Fig. \ref{fig4} demonstrates that in general the reddening law by
\citet{C89} with a fixed $R(V)$ may not reproduce at the same time 
FUV and NUV magnitudes of CSFGs, requiring systematically lower $R(V)$
at shorter wavelengths. Therefore, the reddening law in CSFGs 
may deviate from that of \citet{C89}.

It was shown by \citet{C94} that empirical curves for SFGs completely lack the
2175\AA\ feature. Later, \citet{C04} and \citet{G04} also found that
empirical reddening curves for AGNs lack this feature. Therefore, we examine
this possibility for our CSFGs, removing this feature from \citet{C89} reddening
curves and replacing it by a simple linear interpolation between wavelenghts
1800\AA\ and 2500\AA. This modification would result in lower extinction in the
NIR range and thus brighter $m$(NUV)$_{\rm mod}$. However, on average, the effect
for our CSFGs is very small, not exceeding 0.1 mag compared to the case with
the 2175\AA\ feature.

The even distribution of the differences between the modelled and
observed apparent magnitudes around zero value 
in Figs. \ref{fig2} and \ref{fig4} with
dispersion of $\sim$ 0.7 mag suggests that the main source of the data spread
are the extinction uncertainties of $\sim$1 mag in
the FUV and NUV ranges, which translate to uncertainties of 
$\leq$ 0.1 -- 0.2 mag in the $V$-band extinction $A(V)$ 
and correspondingly $\leq$ 0.1 in $C$(H$\beta$). This is the typical 
uncertainty of the extinction coefficient determination from 
the hydrogen Balmer decrement. Furthermore, the errors of {\sl GALEX} magnitudes
for our faint CSFGs are relatively high and can reach up to $\sim$ 0.3 -- 0.5
mags. This further increases the dispersion in Fig. \ref{fig4}.

In the above discussion we assumed that the extinction for nebular and stellar
radiation are equal. This assumption may not be valid. In particular, it was
argued by e.g. \citet{C00} that the ratio of the stellar extinction 
to the nebular extinction for local SFGs is only 0.44 and may vary in the 
range 0.44 -- 1.0 for both the local and high-$z$ SFGs 
\citep{C97,E06,R10,K13,Pr14,P15,V15,P16}. Low values of the ratio imply that
stars and ionised gas are not co-spatial. In particular, this appearance
is expected for the stars located in the holes which are surrounded by denser
ionised gas clouds. High ratios are in favour of more uniform gas 
distributions around the stars.

We address this issue for our galaxies assuming that the nebular extinction 
derived from the Balmer decrement is two times higher than the stellar 
extinction, but reddening laws are the same. 
Results are presented in Fig.~\ref{fig5}a for the FUV range and
in Fig.~\ref{fig5}b for the NUV range assuming the \citet{C94} reddening law. 
The effect is higher for the FUV
magnitudes because of the higher extinction, but in both diagrams modelled 
brightnesses are overpredicted with a clear trend toward higher intrinsic 
galaxy luminosities, while differences in 
$m$(FUV)$_{\rm mod}$ -- $m$(FUV)$_{\rm obs}$ and 
$m$(NUV)$_{\rm mod}$ -- $m$(NUV)$_{\rm obs}$ are much lower for the 
case with equal nebular and stellar extinctions 
(Figs.~\ref{fig2}d and \ref{fig4}d). 
The conclusions are not changed if dependencies of 
$m$(FUV)$_{\rm mod}$ -- $m$(FUV)$_{\rm obs}$ and 
$m$(NUV)$_{\rm mod}$ -- $m$(NUV)$_{\rm obs}$ on SFR or $M_\star$ are considered, 
because these integrated characteristics correlate with $M$(FUV) and $M$(NUV).
This comparison favours small differences 
between the nebular and stellar extinctions.

\begin{figure}
\includegraphics[angle=-90,width=0.85\linewidth]{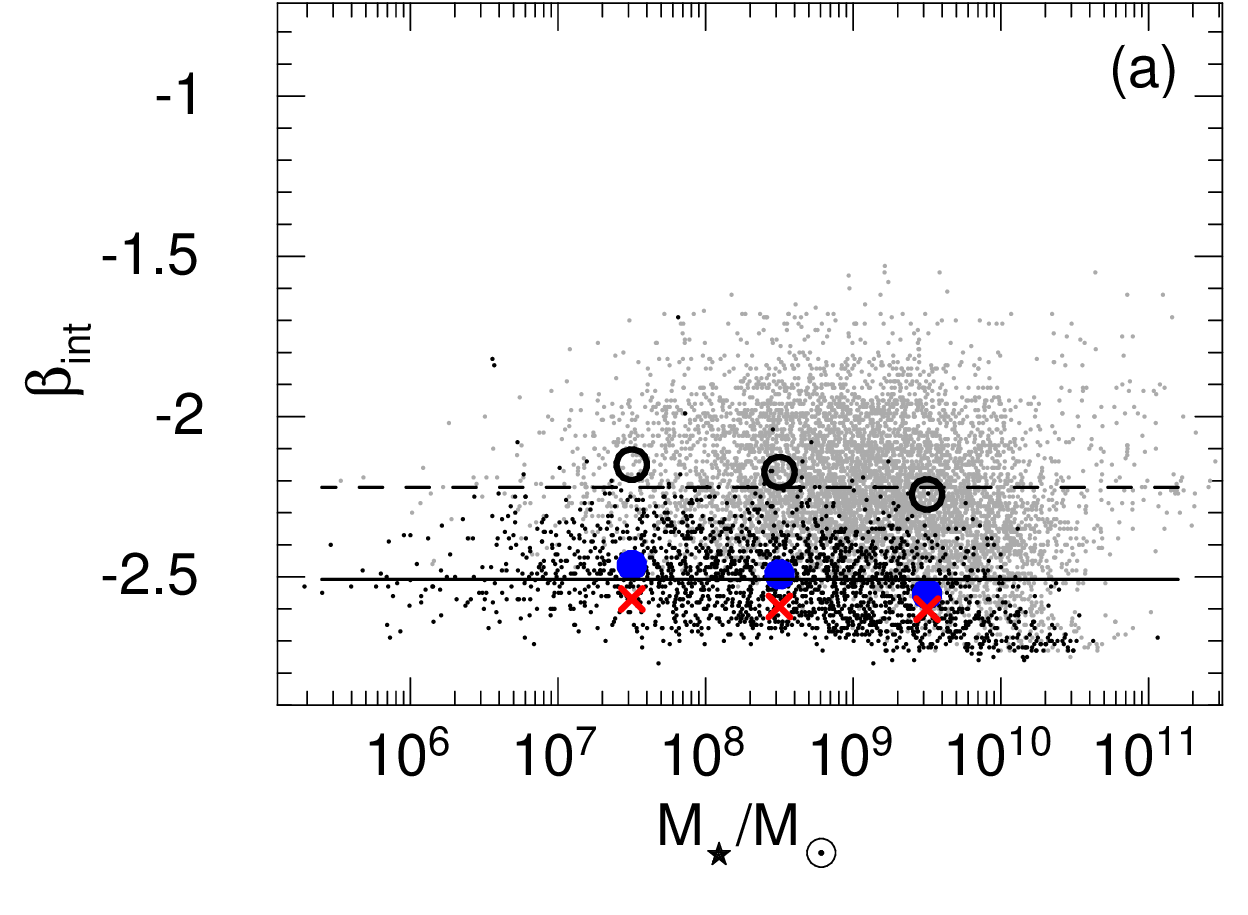}
\includegraphics[angle=-90,width=0.85\linewidth]{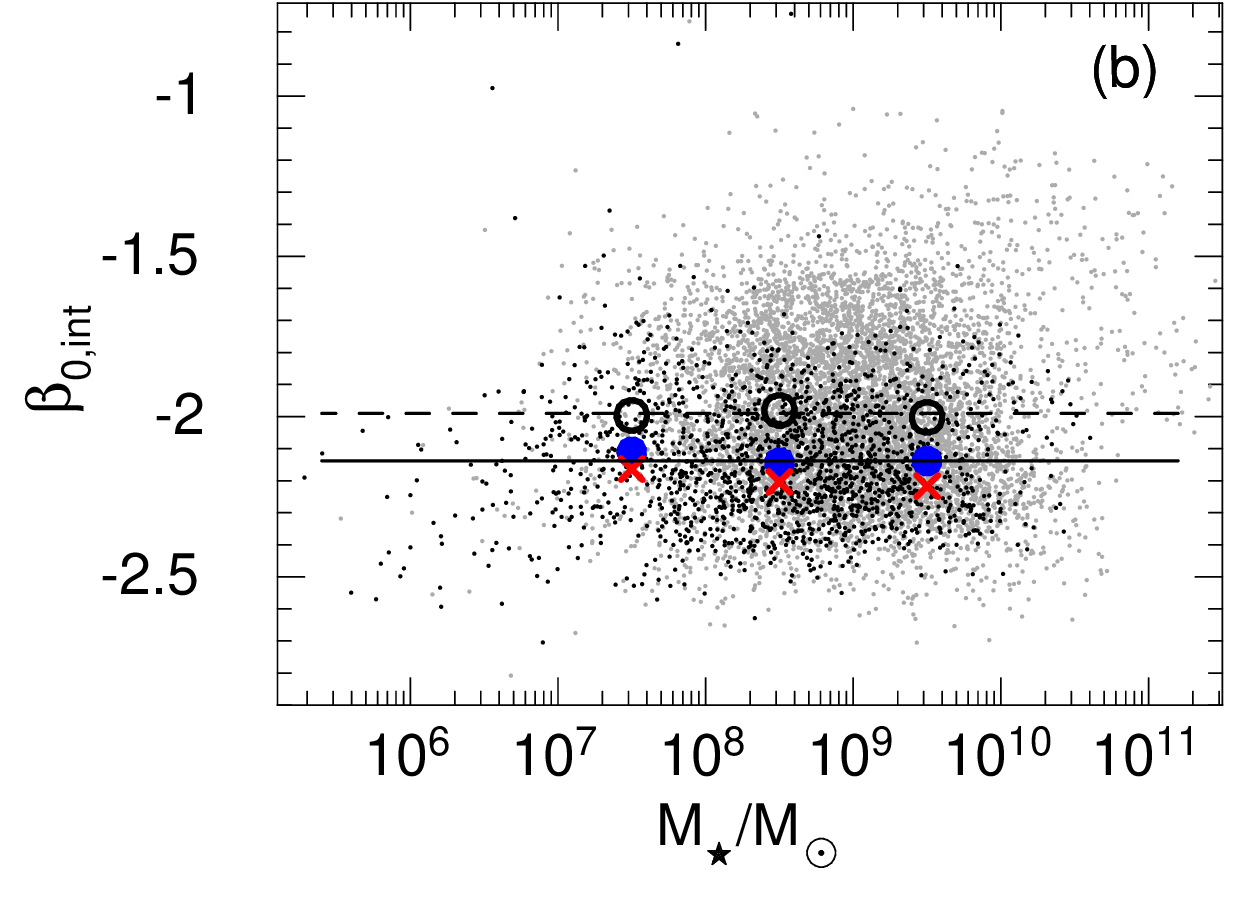}
\includegraphics[angle=-90,width=0.85\linewidth]{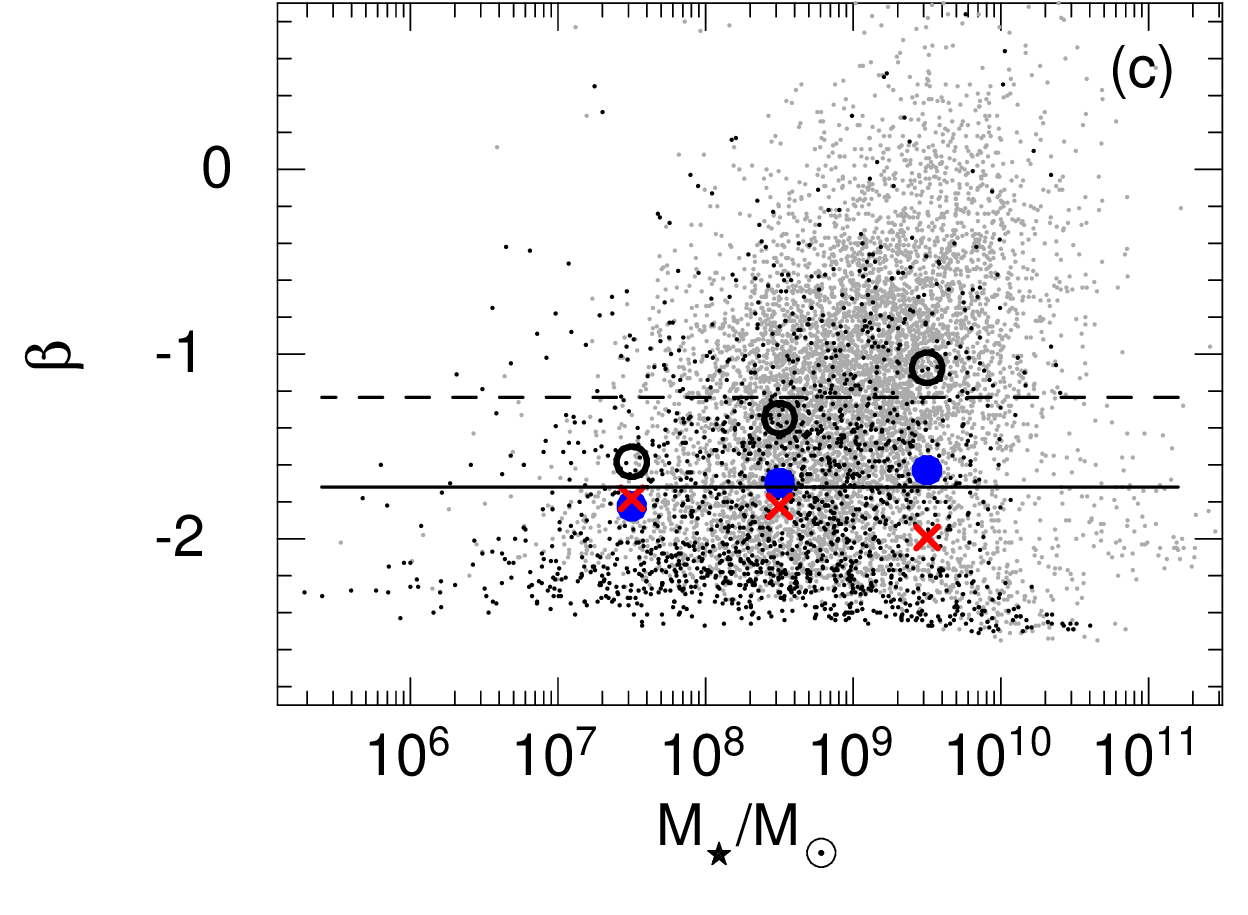}
\caption{The dependence of (a) the intrinsic 
UV continuum slope $\beta_{\rm int}$ and (b) the intrinsic UV continuum 
slope $\beta_{\rm 0,int}$, reduced to zero starburst age, on the stellar 
mass of the galaxies. (c) Same as in 
(a) and (b), but UV continuum slopes $\beta$ are derived from obscured 
SEDs with extinction coefficient $C$(H$\beta$) and adopting $R(V)$ = 2.7 for 
galaxies with 
EW(H$\beta$) $\geq$ 50\AA\ (black dots) and $R(V)$ = 3.1 for galaxies with 
EW(H$\beta$) $<$ 50\AA\ (grey dots). Solid and dashed horizontal lines in
all panels are average values of $\beta$ in CSFGs with high and low 
EW(H$\beta$), respectively. The average values in 1 dex bins of log $M_\star$ 
for CSFGs with
EW(H$\beta$) $\geq$ 150\AA, $\geq$ 50\AA, and $<$ 50\AA\ are shown 
by red crosses, blue filled circles and black open circles, respectively.
\label{fig6}}
\end{figure}

\section{The slope of the UV continuum in CSFGs}\label{sec4}

The shape of the UV continuum is widely used to study extinction in
low-$z$ and high-$z$ SFGs. In particular, the empirical \citet{C94} 
reddening law is parameterized by the slope $\beta$ of the UV continuum
adopting a power law flux distribution $I(\lambda)$ $\sim$ $\lambda^\beta$. We 
define the slopes $\beta_{\rm int}$ and $\beta$ of the intrinsic and obscured 
SEDs, respectively:
\begin{equation}
\beta_{\rm int} = \frac{\log I(\lambda_1) - \log I(\lambda_2)}{\log \lambda_1 - \log \lambda_2}, \label{beta_int}
\end{equation}
\begin{equation}
\beta = \beta_{\rm int} - 0.4\times\frac{A(\lambda_1) - A(\lambda_2)}{\log \lambda_1 - \log \lambda_2}, \label{beta_red}
\end{equation}
where $\lambda_1$=1300\AA\ and $\lambda_2$=1800\AA\ are rest-frame
wavelengths, $I_\lambda$ is intrinsic flux, which includes both the stellar and
nebular emission, and $A(\lambda)$ is the extinction in mags.

In Fig. \ref{fig6}a we show the dependence of the intrinsic slope 
$\beta_{\rm int}$ on
the stellar masses $M_\star$ of the $\sim$ 14000 galaxies from our sample. 
As in Figs. \ref{fig2} -- \ref{fig5},
the sample is split into CSFGs with low EW(H$\beta$) (grey dots) and
high EW(H$\beta$) (black dots). 
We note that for galaxies whose optical light is dominated by young stars,
as the case for most sources considered here, EW(H$\beta$) is primarily an
age indicator. For other galaxies EW(H$\beta$) will be more sensitive
to other parameters as the specific SFR.

A clear separation between these two 
CSFG subsamples is present with steeper slopes 
of $\sim$ --2.3 to --2.7 for galaxies with high EW(H$\beta$), while slopes
of the UV continuum in galaxies with EW(H$\beta$) $<$ 50\AA\ are
$\sim$ --1.8 to --2.5. This is expected because star-forming regions in CSFGs 
with high EW(H$\beta$) from our sample have younger ages and the radiation of 
short bursts of SF is dominant in the UV range \citep{I16c}.

Adopting the starbursting scenario
we apply the correction of the UV slope to reduce it to zero burst age
assuming that SF occurs in a single instantaneous burst and
using Starburst99 models \citep{L99,L14}. This correction includes both the
stellar and nebular emission which both depend on the galaxy metallicity
and the starburst age. Stellar emission is reduced to zero age adopting
the dependencies of intrinsic fluxes at $\lambda_1$=1300\AA\ and 
$\lambda_2$=1800\AA\ on EW(H$\beta$) in Starburst99 instantaneous burst models
which depend also on the metallicity. As for nebular continuum at the same
wavelengths, it was scaled up by the factor of 
EW(H$\beta$)$_{t=0}$/EW(H$\beta$)$_{\rm obs}$, where EW(H$\beta$)$_{\rm obs}$ and 
EW(H$\beta$)$_{t=0}$ are H$\beta$ equivalent widths observed and at the burst age
$t=0$, respectively. Finally, age-corrected slopes are derived from the sums of 
stellar and nebular emission at $\lambda_1$=1300\AA\ and $\lambda_2$=1800\AA\
using Eq. \ref{beta_int}.

We present in Fig. \ref{fig6}b the dependence of the intrinsic UV slope 
$\beta_{\rm 0,int}$ reduced to a zero age on the galaxy stellar mass. Comparing 
with Fig. \ref{fig6}a we conclude that the age correction makes the slope 
flatter with average values shown by solid and dashed lines. The flattening
is caused by the fact that the slope of the intrinsic stellar emission
of the young starburst corresponding to $\beta$ $\sim$ --2.8 is much steeper 
than the nebular emission with a slope corresponding to $\beta$ $\sim$ 0.0. 
The fraction of nebular emission for younger ages is higher making the slope of
total stellar and nebular emission flatter.
The correction for the burst age considerably reduces the offset between CSFGs 
with high and low EW(H$\beta$)s indicated by the difference of average 
$\beta_{\rm 0,int}$ values (solid and dashed lines) confirming
conclusions made earlier, e.g., by \cite{I11,I15,I16c} that SF
in these galaxies mainly has a bursting nature.

Fig. \ref{fig6}c shows the dependence of 
the UV slopes $\beta$ on stellar masses $M_\star$. $\beta$ has been  
derived from the obscured SEDs adopting extinction obtained from the Balmer
decrement and reddening laws with $R(V)$ = 2.7
for galaxies with EW(H$\beta$) $\geq$ 50\AA\ and with $R(V)$ = 3.1
for galaxies with EW(H$\beta$) $<$ 50\AA. Stellar and interstellar extinction 
were assumed to be the same. The spread of slopes in this panel 
is much higher than in Figs. \ref{fig6}a,b 
implying that it is mainly determined by extinction, which is responsible
for $>$ 70\% of the spread, rather than by different 
starburst ages. The range of slopes $\beta$ for our CSFGs is --2.5 - 0.6,
similar to that for high-redshift UV-selected SFGs with $z$ = 2 - 10
\citep{Fi12,B14,B16}. Average values of $\beta$ shown in 
Fig. \ref{fig6}c by horizontal lines and various symbols are comparable to those
for $z$ $\sim$ 4 galaxies \citep{Fi12} and for $z$ $>$ 1 galaxies 
\citep{R10,R15,Ku14,F16,M16}.

\begin{figure}
\includegraphics[angle=-90,width=0.85\linewidth]{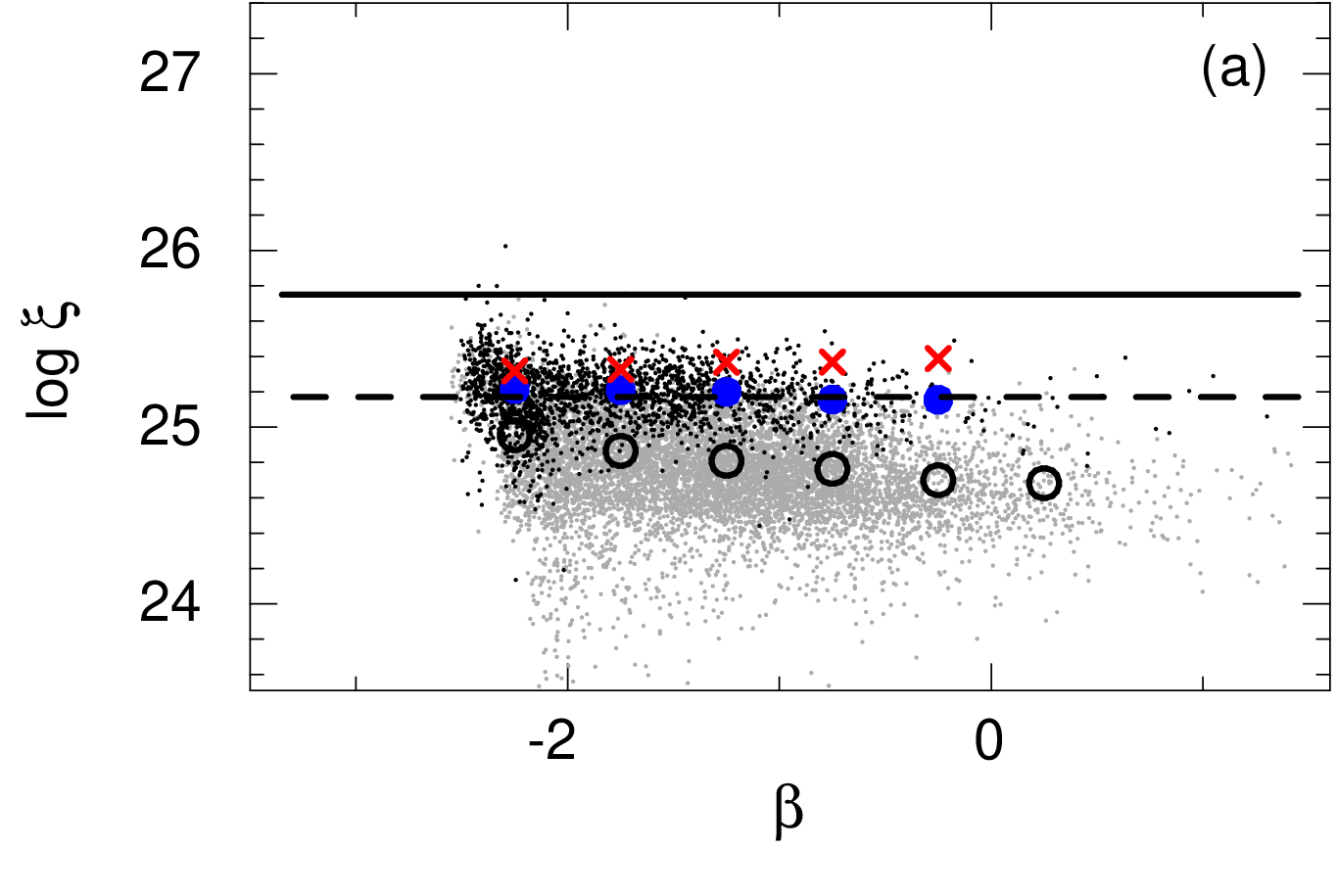}
\includegraphics[angle=-90,width=0.85\linewidth]{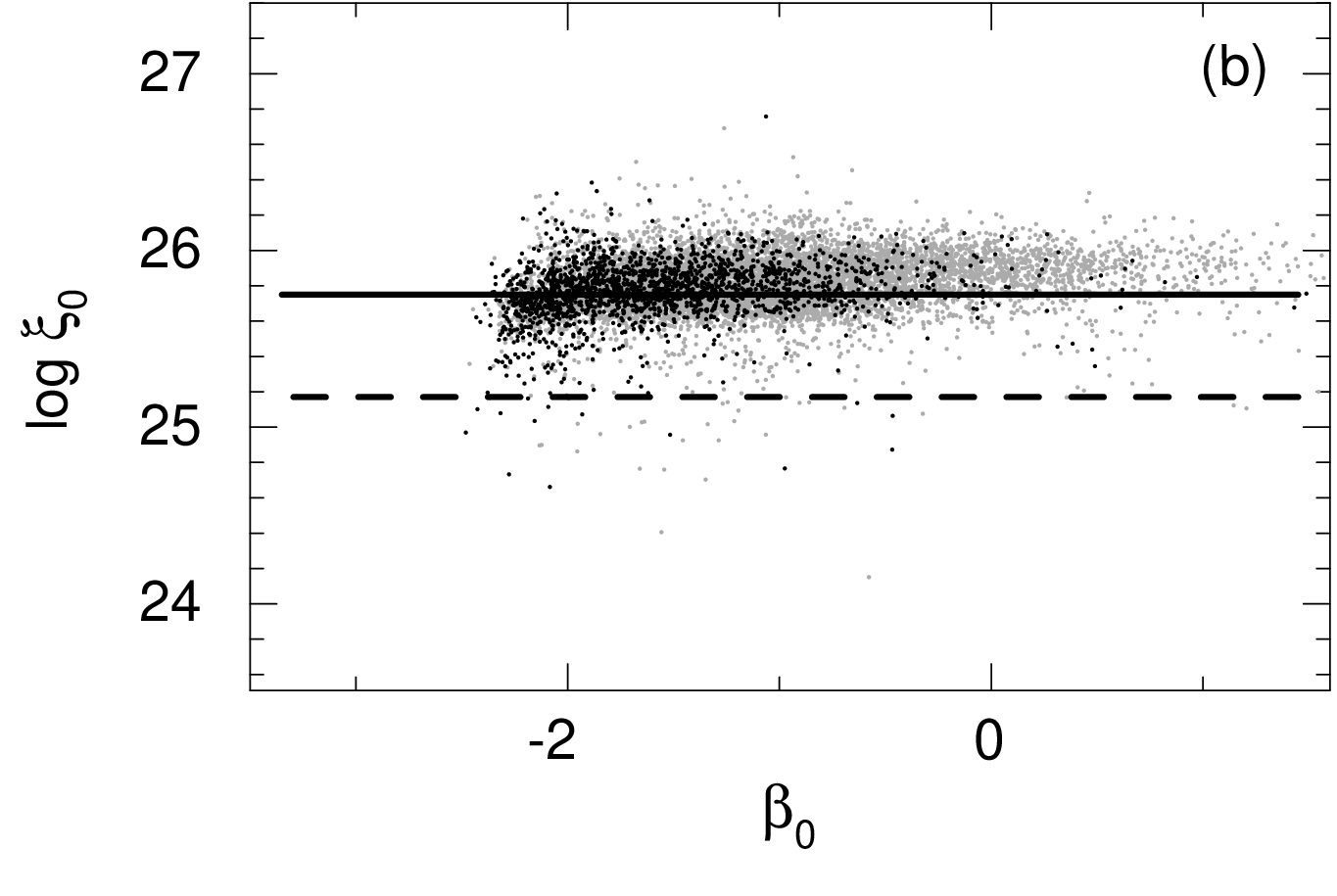}
\includegraphics[angle=-90,width=0.85\linewidth]{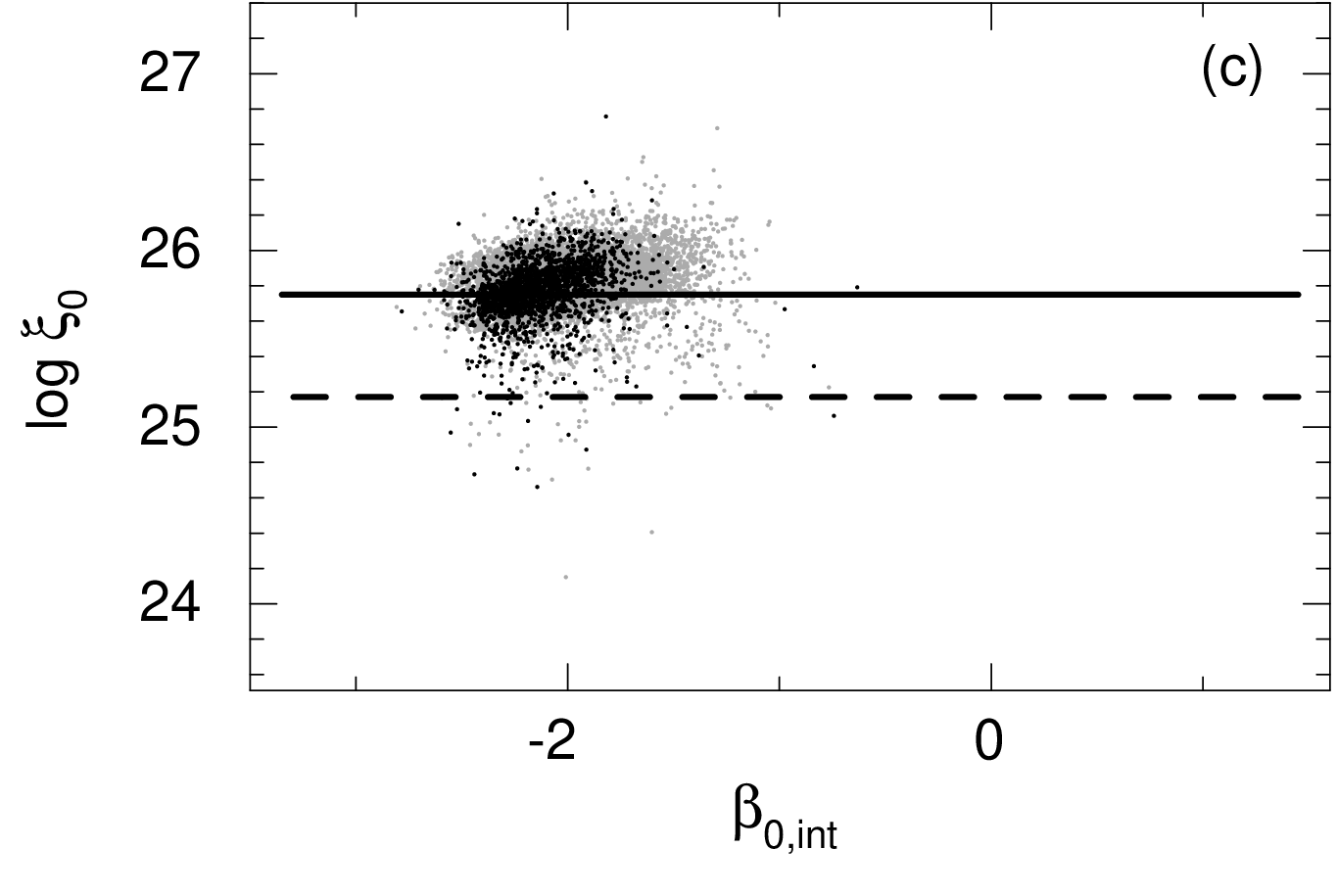}
\caption{(a) The dependence of the efficiency of ionising photon 
production $\xi$ 
on the UV continuum slope $\beta$ of the obscured SEDs with the extinction 
derived from the hydrogen Balmer decrement in the SDSS spectra and adopting 
$R(V)$ = 2.7 for galaxies with EW(H$\beta$) $\geq$ 50\AA\ (black dots) and 
$R(V)$ = 3.1 for galaxies with EW(H$\beta$) $<$ 50\AA\ (grey dots),
while blue filled and black open circles represent respective 
log~$\overline{\xi}$ in 0.5 bins of $\beta$, where $\overline{\xi}$
are average $\xi$ values. Additionally, 
log~$\overline{\xi}$ for CSFGs with 
EW(H$\beta$) $\geq$ 150\AA\ are shown by red crosses.
(b) same as in (a) but $\xi_{\rm 0}$ (the efficiency of ionising photon 
production) and $\beta_0$ (the obscured continuum UV slope) are both reduced to 
zero age. 
(c) same as in (b) but $\beta_{\rm 0,int}$ is the slope of the intrinsic UV 
continuum. 
Solid and dashed horizontal 
lines in all panels indicate ionising photon production for zero-age 
instantaneous bursts and for continuous SF, respectively. 
\label{fig7}}
\end{figure}

\begin{figure}
\includegraphics[angle=-90,width=0.85\linewidth]{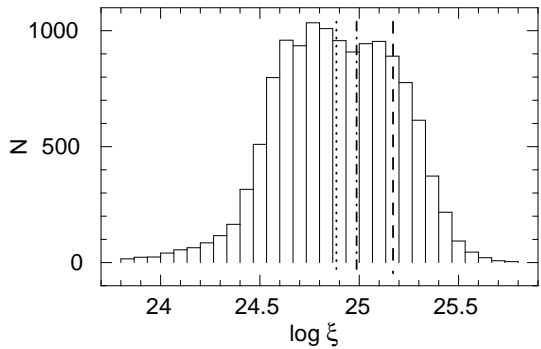}
\caption{The histogram of the efficiency of ionising photon production 
$\xi$ 
distribution for
the entire CSFG sample. Dotted and dash-dotted vertical 
lines indicate $\overline{\log \xi}$ and
log~$\overline{\xi}$, which are the average value of log $\xi$ and 
the log of the average value of $\xi$, respectively. The value
of log~$\xi$ for the model with continuous star formation is shown by the 
dashed vertical line.
\label{fig8}}
\end{figure}

\section{Efficiency of ionising photon production}\label{sec5}

We noted above that CSFGs are characterized by active star formation and have
properties, which in many respects are similar to high-$z$ SFGs \citep{I15}. 
Because our galaxies produce a copious amount of ionising photons they can be
considered as analogs of high-$z$ galaxies responsible for the reionisation of 
the Universe at redshifts $z$~=~5~--~10. One of the important parameters
regulating the intergalactic medium (IGM) ionisation is the LyC escape fraction
$f_{\rm esc}$(LyC). \citet{I16a,I16b} have shown that the absolute
escape fraction $f_{\rm esc}$(LyC) in local
CSFGs can be as high as $\sim$ 10\%. Another parameter is the
efficiency of ionising photon production determined as
\begin{equation}
\xi = \frac{N({\rm LyC})}{L_\nu}, \label{ksi}
\end{equation}
where $N({\rm LyC})$ is the Lyman continuum photon production rate,
$L_\nu$ is the intrinsic monochromatic luminosity at the rest-frame
wavelength of 1500\AA.
$L_\nu$ in our galaxies is derived from the SED fitting, while
the Lyman continuum photon production rate is derived from the relation
using the extinction-corrected H$\beta$ luminosity $L$(H$\beta$)
according to \citet{SH95}:
\begin{equation}
N({\rm LyC}) = 2.1\times 10^{12} L({\rm H}\beta), \label{NLyC}
\end{equation}
where $N$(LyC) and $L$(H$\beta$) are in units s$^{-1}$ and erg s$^{-1}$,
respectively.
Knowing $f_{\rm esc}$(LyC) and $\xi$ it is possible to compute the total photon 
rate at which a galaxy population ionises the IGM \citep[e.g. ][]{R13}.
The production efficiency $\xi$ of a given stellar population
is a simple prediction from synthesis models, which depends
on metallicity, star-formation history, age, and also on assumptions on stellar
evolution. Typically canonical values of log $\xi$ $\sim$ 25.2 -- 25.3
are adopted for high-$z$ studies, where $\xi$ is expressed in erg$^{-1}$Hz. 
Observationally $\xi$ has recently been estimated by
\citet{B15} for a sample of high-$z$ LBGs, finding values of $\xi$ 
compatible with the canonical value. 
However, the ionising photon production of galaxies known to be
LyC leakers (i.e. with $f_{\rm esc}$(LyC) $>$ 0) has not been measured until
recently. First determinations in five low-$z$ LyC leakers were presented by
\citet{S16} who found log $\xi$ $\sim$ 25.1 -- 25.5.

In this paper we derive the production efficiency for all $\sim$~14000 CSFGs 
selected from the SDSS. Fig. \ref{fig7}a shows the dependence of $\xi$ on the 
slope $\beta$ of the obscured SED for CSFGs with
EW(H$\beta$) $\geq$ 50\AA\ (black dots) and EW(H$\beta$) $<$ 50\AA\ 
(grey dots). It is seen that data scatter in a wide range log $\xi$ 
$\sim$ 24.5 -- 25.8 with highest values for CSFGs with highest EW(H$\beta$). 
Black open circles, blue filled circles and red crosses  
show logs of average values $\overline{\xi}$ of $\xi$ in 0.5 bins of $\beta$ 
for CSFGs with EW(H$\beta$) $<$ 50\AA, $\geq$ 50\AA\ and 
$\geq$ 150\AA, respectively. The average values of $\xi$ in log scale are 
log ${\overline{\xi}}$ $\sim$ 24.82, $\sim$ 25.21 and $\sim$ 24.93 for CSFGs
with EW(H$\beta$) $\geq$ 50\AA, with EW(H$\beta$) $<$ 50\AA, and for the entire 
sample, respectively. Selecting only galaxies with 
[O~{\sc iii}]/[O~{\sc ii}] $\ga$ 5
\citep[the values characteristic of LyC escaping galaxies, ][]{I16a,I16b}
we obtain log~$\overline{\xi}$ = 25.31. All these values are similar to the
canonical value implying that CSFGs could be important contributors
of escaping ionising radiation capable to ionise the IGM.
The dispersion of $\xi$ is likely not dominated by uncertainties in the dust 
corrections. Although these uncertainties might be relatively high with
dispersion of $\sim$ 0.7 mag in the FUV range or 0.3 dex in $L$(1500\AA) 
(see Sect. \ref{sec3}),
they can not produce an offset for CSFGs with high EW(H$\beta$) relative 
to the galaxies with low EW(H$\beta$). IMF variations can be present but this 
is a highly speculative and uncertain topic, not warranted by observations.

A similar range of log $\xi$ is found for high-$z$ LBGs and Ly$\alpha$ emiters
(LAEs) with $z$ $\sim$ 3 -- 7.7 \citep{B15,N16,St16}. 
Comparing our values of $\xi$ for galaxies
with high EW(H$\beta$) with those for $z$ $\sim$ 4 -- 5 galaxies
by \citet{B15b} we find good agreement. The range of
log $\xi$ for our galaxies (black dots) is also consistent with the
results of modelling by \citet{Wi16b} for different stellar population
synthesis models. However, log $\xi$ of our entire sample of
CSFGs is on average higher than the average value of 24.77 for $z$ = 2.2 
H$\alpha$ emitting galaxies by \citet{M16}. 

The distribution of $\xi$s
for CSFGs with the lowest EW(H$\beta$) (grey dots in Fig. \ref{fig7}a) shows a 
trend of decreasing $\xi$ with increasing $\beta$. On the other hand, 
no such trend is detected for CSFGs with higher EW(H$\beta$)s. 

For comparison, solid and dashed lines in Fig. \ref{fig7}a
represent the values of log $\xi$ for the
Starburst99 zero-age instantaneous burst models and the models with 
continuous SF with constant SFR, respectively. They take into account the 
contribution of both the stellar and nebular emission at 1500\AA\ and are
consistent with the respective values derived by \citet*{Re10} for 
heavy element mass fraction greater than 10$^{-2}$.

A histogram of the $\xi$ distribution for the entire sample of CSFGs
is shown in Fig. \ref{fig8}. The average value $\overline{\xi}$ of $\xi$ shown 
by dash-dotted line is $\la$ 1.5 times lower than the value for the model
with continuous SF (dashed line). However, the average values of $\xi$ for CSFGs
with EW(H$\beta$) $\geq$ 50\AA\ (blue filled circles in Fig. \ref{fig7}a) are 
very similar to the value for the model with continuous SF.

It is clear from Fig. \ref{fig7}a that the model with continuous SF 
fails to reproduce observations both for CSFGs with high and low EW(H$\beta$), 
which scatter in a wide range around the dashed line. This model predicts nearly constant 
EW(H$\beta$) $\sim$ 40 -- 50\AA\ and constant $\xi$,
while the range of EW(H$\beta$) for galaxies from our sample is 
$\sim$10--400\AA\ and $\xi$ varies by one order of magnitude.
On the other hand, the models with the instantaneous bursts 
may be succesful in reproducing observations. In this case $\xi$ should
rapidly vary with age of the burst. The bursting nature of star formation
is a reason why distributions of CSFGs with high and low EW(H$\beta$)
are different in Fig. \ref{fig7}a. To show this we need to reduce
the efficiency of ionising photon production to zero burst age, which we
denote as $\xi_0$.
For this we use Starburst99 models with appropriate metallicities
\citep{L99,L14} taking into account both the stellar and nebular emission in 
the continuum at 1500\AA.

As it could be seen in Fig. \ref{fig7}b, distributions of $\xi_0$ for CSFGs 
with high and low EW(H$\beta$)s are similar. There is a small upward offset
of CSFGs with low EW(H$\beta$) (grey dots) relative to the 
value of the zero-age instantaneous burst model (solid line).
This can be accounted for by the non-negligible 
contribution of an older stellar population to the continuum near the H$\beta$
emission line, which is higher for galaxies with lower EW(H$\beta$).
As a result, the age correction would be overestimated if the instantaneous 
burst model is taken. However, since the upward offset is only $\leq$ 0.1 dex,
the assumption of an instantaneous burst model is reasonable. This assumption
is further supported by Fig. \ref{fig7}c, where the distribution 
of $\xi_0$ on the intrinsic UV slope reduced to zero age, $\beta_{0,{\rm int}}$,
is shown. Both $\xi_0$ and $\beta_{0,{\rm int}}$
occupy a small region with dispersions, which can be caused by 
uncertainties of extinction determination, SED fitting and reduction to zero
burst age, but in general the distribution of CSFGs is well reproduced 
by the instantaneous burst models.

We note, however, that the assumption of a continuous SF with 
constant SFR could be a reasonable approximation in estimating the average 
global efficiency of ionising 
photon production, e. g., in the early Universe during the epoch of 
reionisation if it is assumed that during that epoch first galaxies were
formed with a constant rate, despite the fact that SF in individual
galaxies proceeded during short bursts. Thus, the value log $\xi$ = 25.17 shown
in Fig. \ref{fig7} by the dashed line, similar to estimates
in earlier studies \citep[e.g. ][]{R13}, is a reasonable estimate, which could
be used in studies of the reionisation of the Universe. However, it may be 
somewhat higher in the early Universe because of lower stellar metallicities 
and a more top-heavy IMF of stars \citep[e.g. ][]{Re10}.

\begin{figure}
\includegraphics[angle=-90,width=0.90\linewidth]{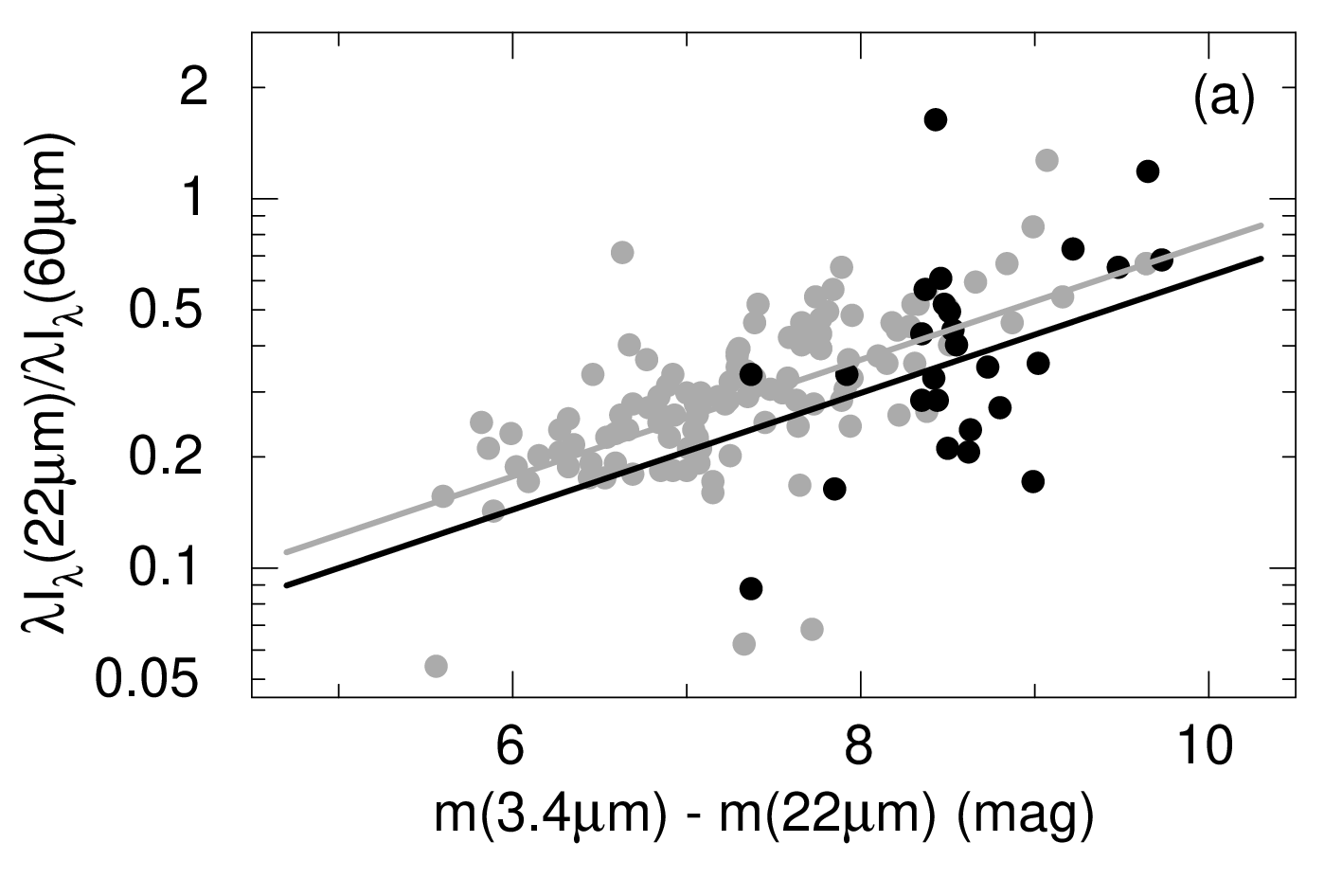}
\includegraphics[angle=-90,width=0.90\linewidth]{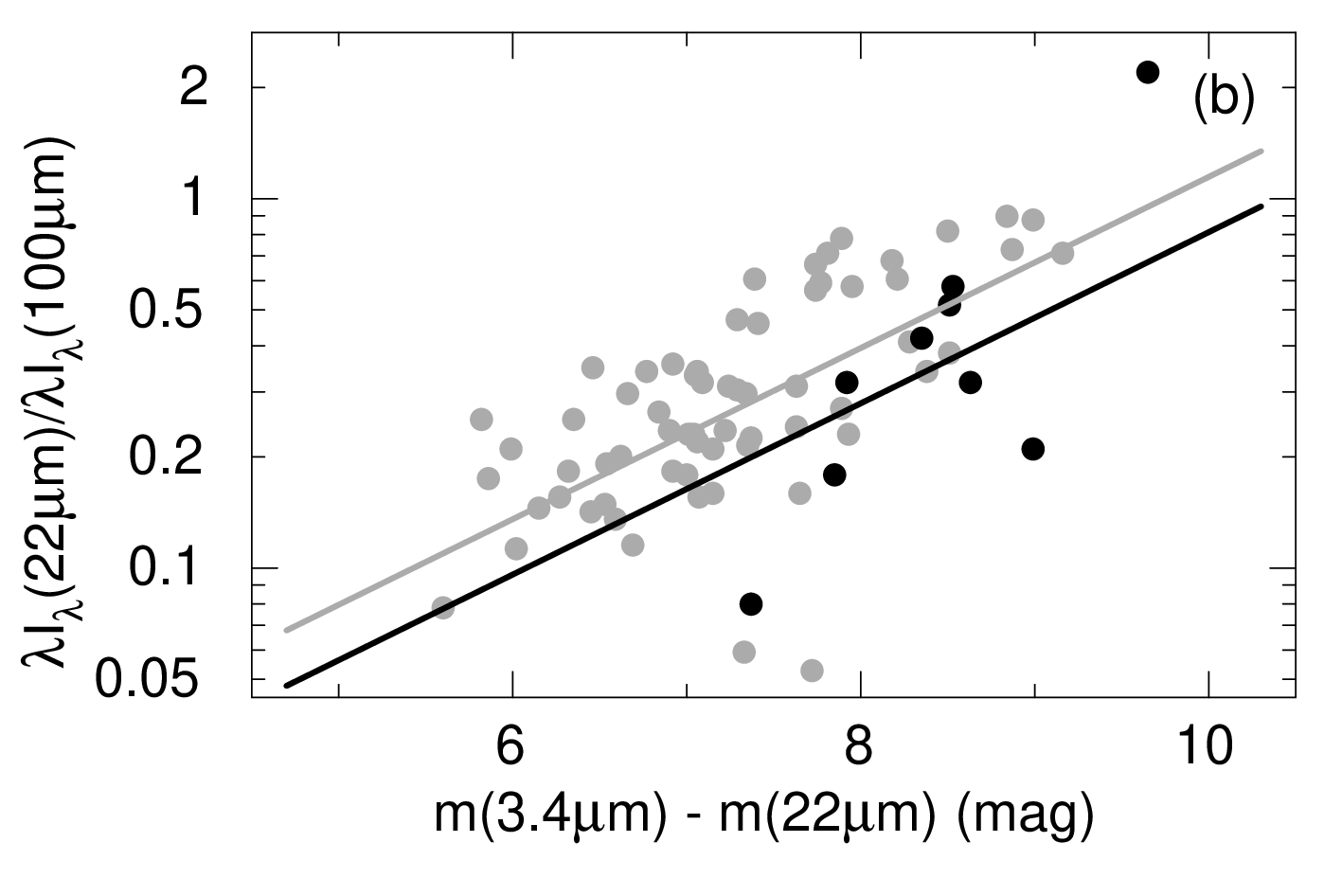}
\caption{The relations between (a) the 22$\mu$m/60$\mu$m intensity 
ratio and (b) the 22$\mu$m/100$\mu$m intensity ratio on the {\sl WISE}
3.4$\mu$m -- 22$\mu$m colour for the galaxies from our SDSS sample, which
were detected by {\sl IRAS} at 60$\mu$m and 100$\mu$m, respectively.
The galaxies with EW(H$\beta$)~$\geq$~50\AA\ and EW(H$\beta$)~$<$~50\AA\
are shown by black and grey filled circles, respectively.
Grey lines are maximum likelihood regressions for galaxies with
EW(H$\beta$)~$<$~50\AA, while black lines are relations adopted for
galaxies with EW(H$\beta$)~$\geq$~50\AA.
\label{fig9}}
\end{figure}

\begin{figure*}
\hbox{
\includegraphics[angle=-90,width=0.45\linewidth]{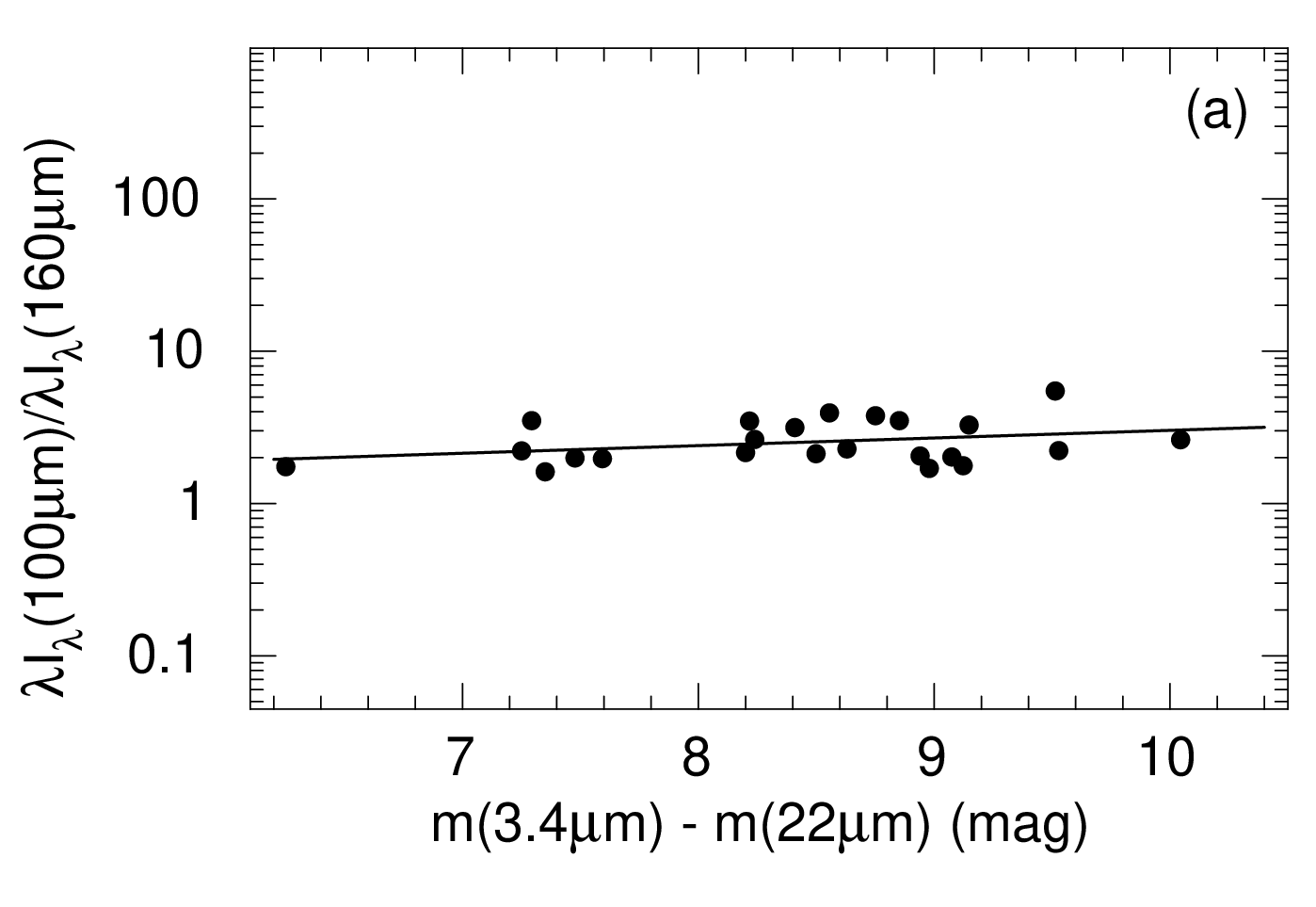}
\hspace{0.3cm}\includegraphics[angle=-90,width=0.45\linewidth]{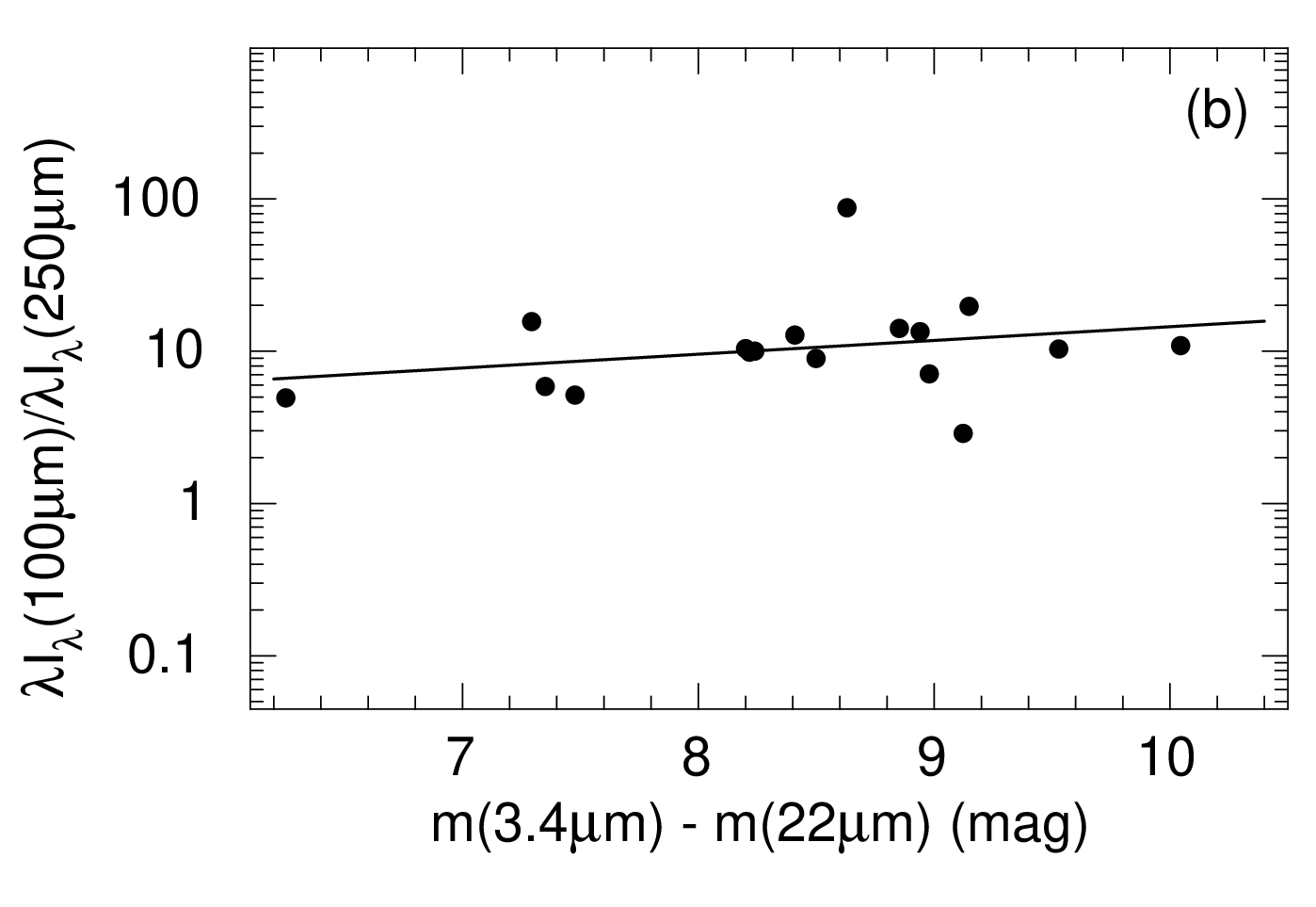}
}
\hbox{
\includegraphics[angle=-90,width=0.45\linewidth]{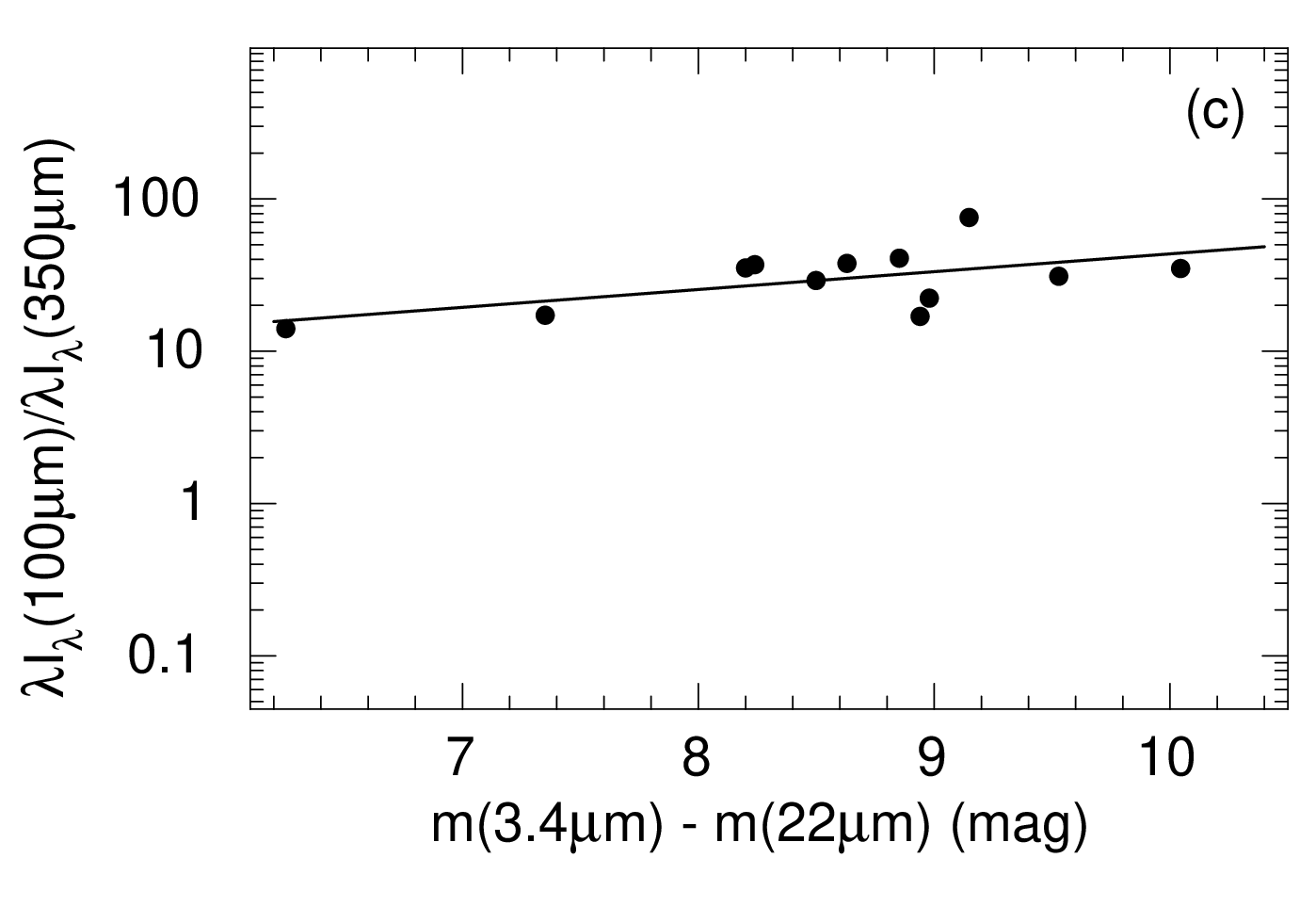}
\hspace{0.3cm}\includegraphics[angle=-90,width=0.45\linewidth]{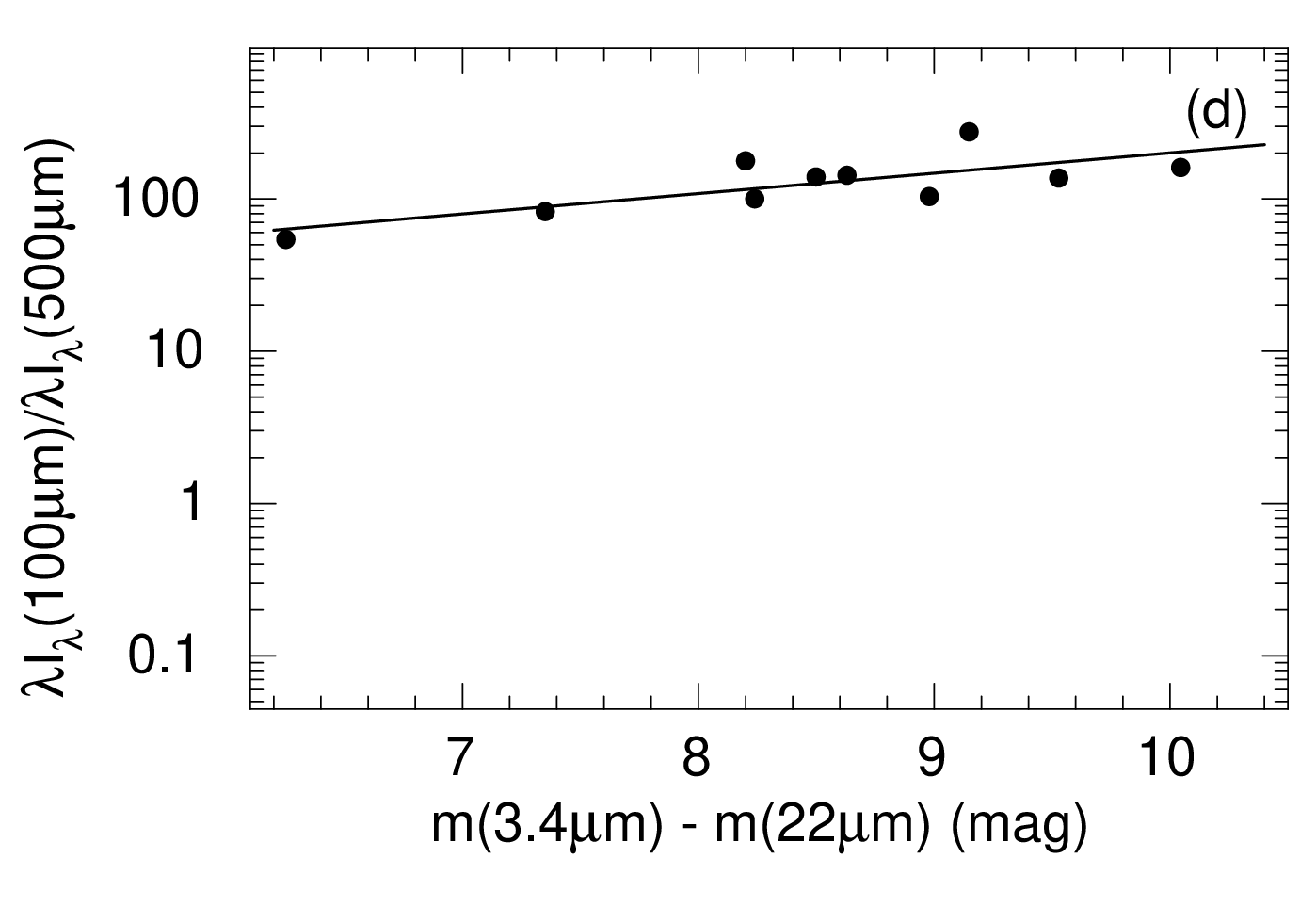}
}
\caption{
The relations between (a) the 100$\mu$m/160$\mu$m intensity 
ratio, (b) the 100$\mu$m/250$\mu$m intensity ratio,
(c) the 100$\mu$m/350$\mu$m intensity ratio and (d) the 100$\mu$m/500$\mu$m 
intensity ratio on the {\sl WISE} 3.4$\mu$m -- 22$\mu$m colour for the 
CSFGs by \citet{RR13}, which
were detected by {\sl Herschel} at 100$\mu$m, 160$\mu$m, 250$\mu$m, 350$\mu$m, 
and 500$\mu$m, respectively. Solid lines are maximum likelihood regressions.
\label{fig10}}
\end{figure*}

\section{The radiation energy balance in CSFGs}\label{sec6}

\subsection{Luminosities in the UV range}

To study the radiation energy balance of the CSFGs we derive the 
energy emitted by the galaxy in the UV and optical wavelength range at $\lambda$
$<$ 1 $\mu$m using modelled SEDs:
\begin{equation}
L({\rm int})=4\pi D^2\int_0^{1\mu {\rm m}}I_\lambda d\lambda, \label{eq:Lint}
\end{equation}
where $I_\lambda$ is the intrinsic SED flux at the rest-frame wavelength 
$\lambda$ and $D$ is the distance. 

Similarly, the absorbed luminosity is determined as
\begin{eqnarray}
L({\rm abs})=&4\pi D^2\exp{(\sqrt{1-\omega}-1)}~~~~~~~~~~~~ \label{eq:Labs} \\
&\times\int_0^{1\mu {\rm m}}I_\lambda [1 - 10^{-(1+f_\lambda)C({\rm H}\beta)}]d\lambda, 
\nonumber
\end{eqnarray}
where term exp($\sqrt{1-\omega}-1$) is the fraction of the absorbed radiation
relative to the sum of absorbed and scattered radiation, $\omega$ is the albedo 
of grains. Here we assume that scattering is isotropic \citep{C94} and 
$\omega$ does not depend on the wavelength. The value of $\omega$ in the UV is 
uncertain. \citet{D03} estimated $\omega$ $\sim$ 0.3 - 0.4, while \citet*{H91} 
found it to be $<$ 0.25 and scattering is fairly isotropic. For clarity, we 
adopt $\omega$ = 0.3. The derived value of $L$(abs) in the case of $\omega$ = 0 (no scattering) would be higher by $\sim$ 13\%.
We neglected a small
contribution of the absorbed stellar and nebular radiation at 
$\lambda$ $>$ 1 $\mu$m because of the very low extinction and low $I_\lambda$
at these wavelengths.


\subsection{Emission in the infrared range}

For the determination of the luminosity in the infrared range we selected
those CSFGs, which were detected by {\sl WISE} in all 
four bands at 3.4$\mu$m, 4.6$\mu$m, 12$\mu$m and 22$\mu$m, totalling 
$\sim$ 3000 of our $\sim$ 14000 CSFGs. We use this sample to select those 145 
galaxies, which were detected by {\sl IRAS} at 60$\mu$m and the 74 galaxies 
seen at 100$\mu$m. These numbers are much smaller than the number of CSFGs 
detected in the mid-infrared range by 
{\sl WISE} in all four bands. Therefore, some assumptions need to be made
to derive the IR fluxes in the far-infrared range. For instance, \citet{B16}
assumed a certain average dust temperature in SFGs. The weakness of this
approach is that the dust mass should be defined as well. Additionally,
it was shown e.g. by \citet{H14} and \citet{I14b} that the dust emission
in SFGs can not be reproduced by a model with a single dust temperature.

We adopt another approach and search for a simple recipe to
derive fluxes at $\lambda$ $>$ 22 $\mu$m using {\sl WISE} fluxes and
apparent magnitudes. In Fig. \ref{fig9} we show
dependencies of 22$\mu$m-to-60$\mu$m and 22$\mu$m-to-100$\mu$m flux ratios
on the {\sl WISE} 3.4$\mu$m -- 22$\mu$m colour indices for SDSS CSFGs. 
These dependencies for galaxies with EW(H$\beta$)~$<$~50\AA\
can be approximated by linear relations
\begin{equation}
\log{\frac{\lambda I_\lambda (22\mu {\rm m})}{\lambda I_\lambda (60\mu {\rm m})}} =
0.158 [m(3.4\mu {\rm m}) - m(22\mu {\rm m})] - 1.700 \label{eq:22_60}
\end{equation}

\noindent and

\begin{equation}
\log{\frac{\lambda I_\lambda (22\mu {\rm m})}{\lambda I_\lambda (100\mu {\rm m})}} =
0.232 [m(3.4\mu {\rm m}) - m(22\mu {\rm m})] - 2.258, \label{eq:22_100}
\end{equation}
which are shown in Fig. \ref{fig9} by grey lines. 
There is a slight offset from these relations for galaxies with higher
EW(H$\beta$)~$\geq$ 50\AA. However, the statistics of these galaxies is too low 
to produce reliable regressions. Therefore, we simply shift regressions shown
by grey lines using average offsets by $\sim$ 0.10 in Fig. \ref{fig9}a
and by $\sim$ 0.15 in Fig. \ref{fig9}b to produce relations for galaxies with 
high EW(H$\beta$) $\geq$ 50\AA\ (black lines in Fig. \ref{fig9}).
All these relations are used to derive fluxes at 60$\mu$m and 100$\mu$m for 
the entire sample of CSFGs detected by {\sl WISE} at 3.4$\mu$m and 
22$\mu$m.

For longer wavelengths we use data for CSFGs by \citet{RR13} obtained 
with the {\sl Herschel} telescope at 100$\mu$m, 160$\mu$m, 250$\mu$m, 
350$\mu$m and 500$\mu$m and derive the following relations (Fig. \ref{fig10}):
\begin{equation}
\log{\frac{\lambda I_\lambda (100\mu {\rm m})}{\lambda I_\lambda (160\mu {\rm m})}} =
0.050 [m(3.4\mu {\rm m}) - m(22\mu {\rm m})] - 0.020, \label{eq:100_160}
\end{equation}

\begin{equation}
\log{\frac{\lambda I_\lambda (100\mu {\rm m})}{\lambda I_\lambda (250\mu {\rm m})}} =
0.090 [m(3.4\mu {\rm m}) - m(22\mu {\rm m})] + 0.258, \label{eq:100_250}
\end{equation}

\begin{equation}
\log{\frac{\lambda I_\lambda (100\mu {\rm m})}{\lambda I_\lambda (350\mu {\rm m})}} =
0.117 [m(3.4\mu {\rm m}) - m(22\mu {\rm m})] + 0.468, \label{eq:100_350}
\end{equation}

\begin{equation}
\log{\frac{\lambda I_\lambda (100\mu {\rm m})}{\lambda I_\lambda (500\mu {\rm m})}} =
0.134 [m(3.4\mu {\rm m}) - m(22\mu {\rm m})] + 0.964. \label{eq:100_500}
\end{equation}

Finally, using mid-infrared fluxes
obtained with {\sl WISE} and the approximations presented in
Eqs. \ref{eq:22_60} -- \ref{eq:100_500} we derive luminosities emitted in the
IR range at wavelengths 8~--~1000~$\mu$m:
\begin{equation}
L({\rm IR})=4\pi D^2\int_{8\mu {\rm m}}^{1000\mu {\rm m}}I_\lambda d\lambda. \label{eq:LIR}
\end{equation}

\begin{figure}
\includegraphics[angle=-90,width=0.99\linewidth]{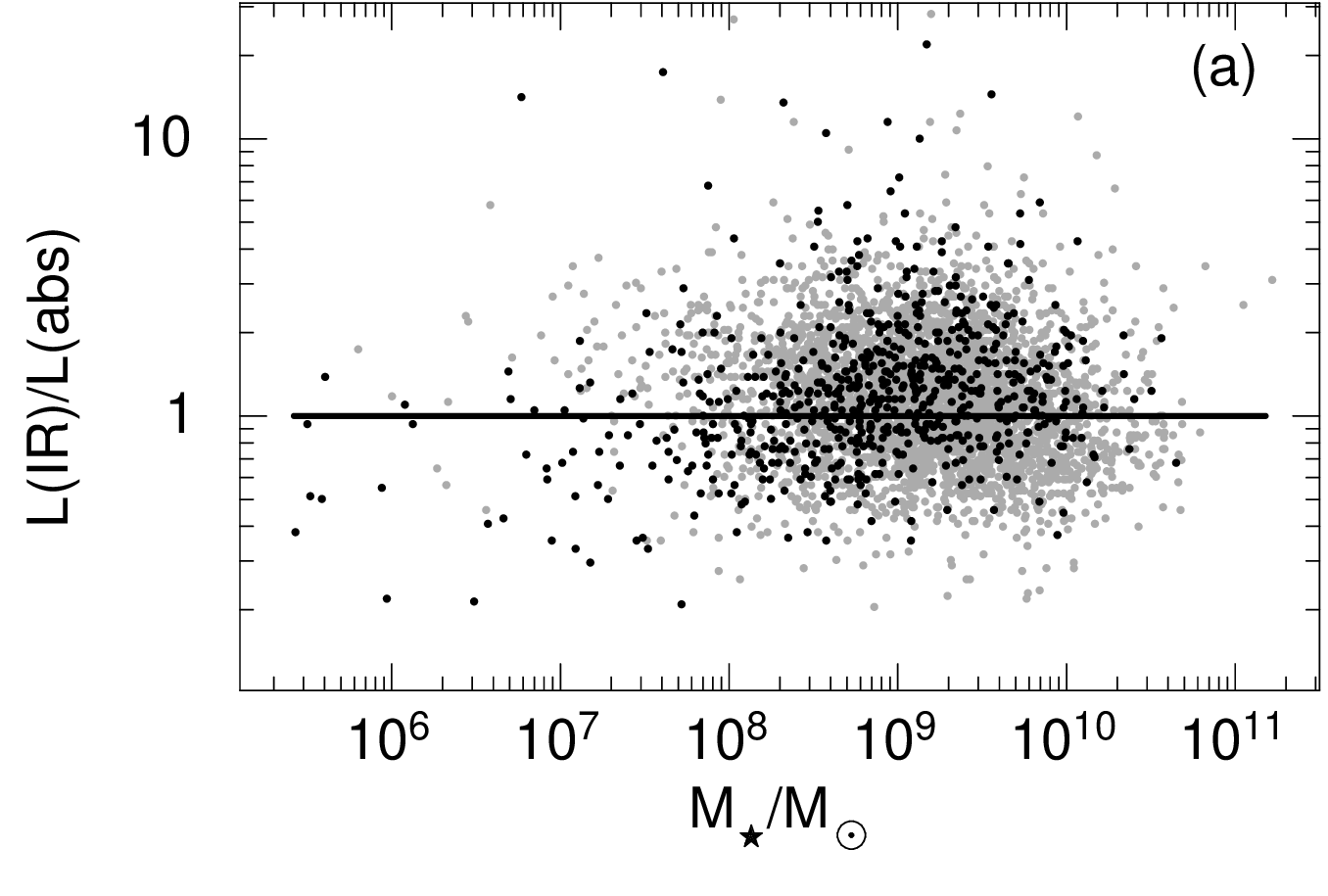}
\includegraphics[angle=-90,width=0.99\linewidth]{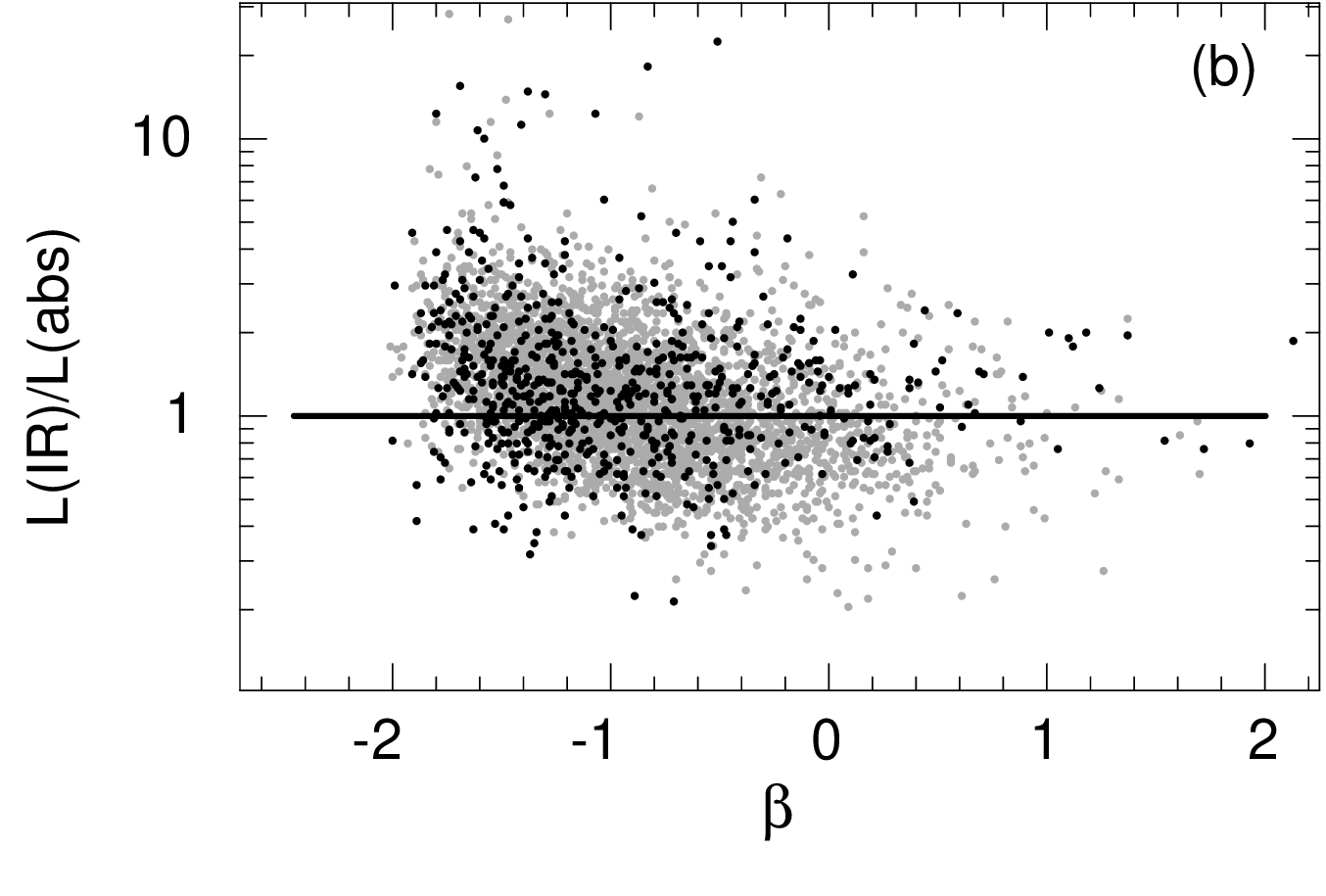}
\caption{The dependencies of the ratios of the infrared luminosities $L$(IR)
in the wavelength range 8 -- 1000 $\mu$m and the luminosities
$L$(abs) absorbed in the wavelength range $<$ 1$\mu$m (a) on
stellar masses and (b) on the obscured spectral UV slope adopting \citet{C89}
extinction laws with $R$($V$) = 2.7 for CSFGs with EW(H$\beta$) $\geq$ 50\AA\
(black dots) and 3.1 for CSFGs with EW(H$\beta$) $<$ 50\AA\ (grey dots) and
extinction coefficients $C$(H$\beta$) which are derived from the hydrogen Balmer
decrement in the SDSS spectra. Solid lines indicate equal $L$(IR) and
$L$(abs). The meaning of symbols is the same as in Fig. \ref{fig2}.
\label{fig11}}
\end{figure}


\begin{figure}
\includegraphics[angle=-90,width=0.99\linewidth]{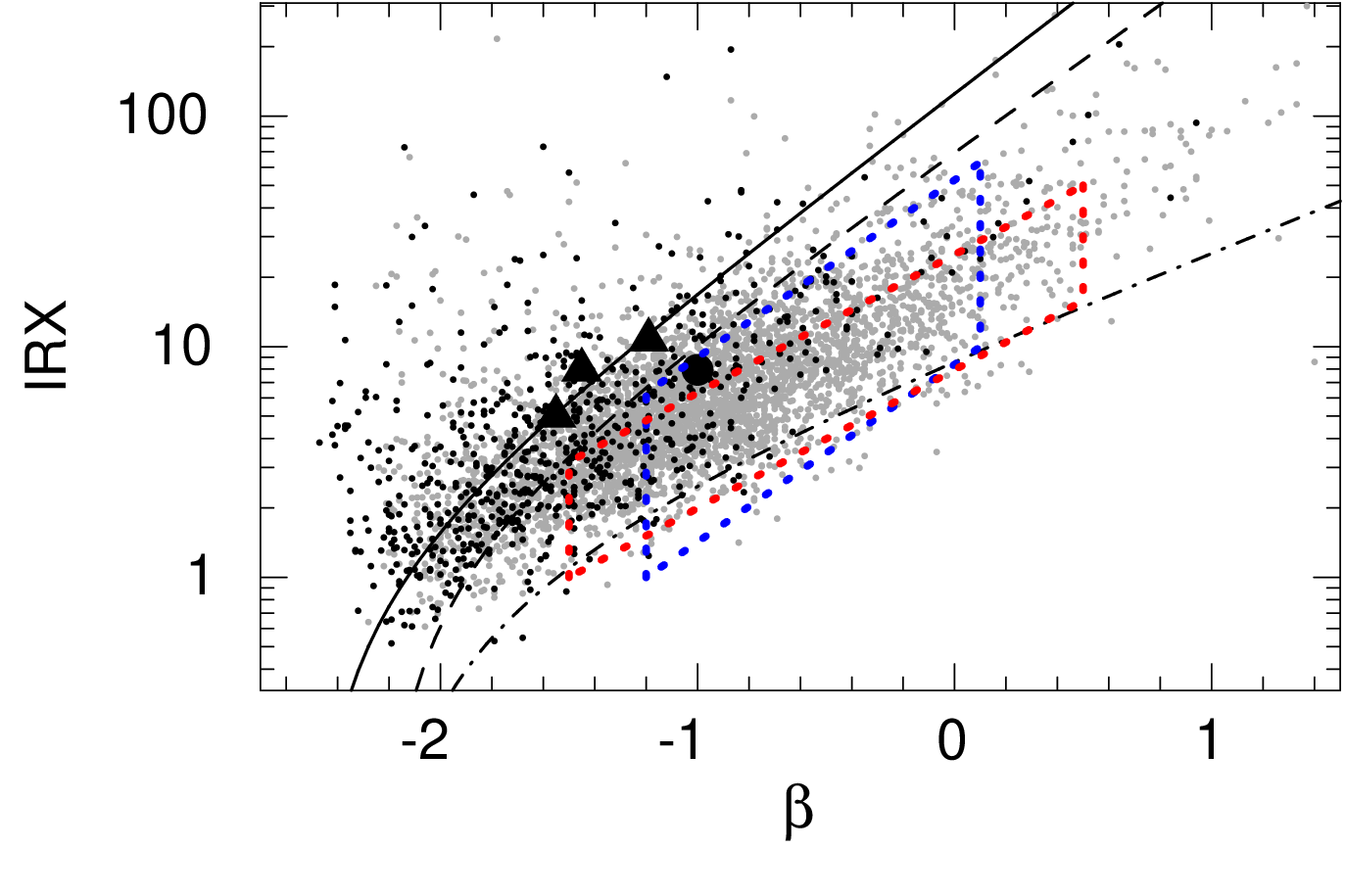}
\caption{The dependencies of the SED infrared excesses $IRX$ 
adopting \citet{C89} reddening laws 
with $R$($V$) = 2.7 for CSFGs with EW(H$\beta$) $\geq$ 50\AA\ (black dots) and
with $R$($V$) = 3.1 for CSFGs with EW(H$\beta$) $<$ 50\AA\ (grey dots)
on the obscured spectral UV slopes $\beta$. Extinctions are 
derived from the hydrogen Balmer decrement in SDSS optical spectra.
The average location of $z$ $\sim$ 2 -- 3 SFGs with stellar masses $>$10$^{9.75}$
M$_\odot$ \citep{B16} and different subsamples of SFGs at $z$ $\sim$ 2 
\citep{R12} are shown by a filled circle and filled triangles, respectively.
The distributions of SFGs by \citet{W16} and \citet{B12} are indicated 
by regions delineated by blue and red dotted lines, respectively.
For comparison, the dependencies for SFGs proposed by \citet{K04} and 
\citet{M99} are shown by solid and dashed lines, while the dependence with
the SMC reddening law by \citet{P98} is shown by a dash-dotted line.
\label{fig12}}
\end{figure}

\subsection{Comparison of the absorbed UV radiation and IR emission}

Using SEDs and Eq. \ref{eq:Labs} we derive the luminosity
absorbed in the wavelength range $\leq$ 1 $\mu$m. We restrict our sample
and include only those CSFGs in which the extinction coefficient 
$C$(H$\beta$) $>$ 0.1 ($\sim$ 4000 galaxies), 
corresponding to an extinction $A(V)$ $\ga$ 0.2 mag. This is done because
the uncertainties of the modelled absorbed luminosities are very high
at low $C$(H$\beta$)s, which are comparable to their uncertainties.
On the other hand, {\sl WISE} fluxes and relations 
Eqs. \ref{eq:22_60} -- \ref{eq:LIR} are used to derive
the luminosity in the infrared range at $\lambda$~=~8~--~1000~$\mu$m. 

Independently derived luminosities $L$(abs) and $L$(IR) are compared
in Fig. \ref{fig11}. It is seen in Fig. \ref{fig11}a that the luminosities 
absorbed at $\lambda$ $\leq$ 1$\mu$m
are nearly equal to the luminosities emitted in the infrared range and do not
depend on the galaxy stellar mass. There is also no dependence on the
age of the starburst as indicated by the absence of an offset
between CSFGs with high and low EW(H$\beta$)s (black and grey dots,
respectively). The $L$(IR)/$L$(abs) ratio is increased
by raising the slope of the obscured UV SED in 
Fig. \ref{fig11}b for $\beta$ from $\sim$ --1 to --2, i.e. with decreasing
extinction. The cause of the trend is not clear. On the other hand, this 
trend is absent at higher $\beta$. 
Perhaps, the trend appears because of the uncertainties in the 
determination of the infrared
luminosities by using relations Eqs. \ref{eq:22_60} -- \ref{eq:LIR}. In any
case, this trend is small and it does not change the general conclusion for the
sample that on average the energy emitted in the IR range is equal to the
energy of radiation absorbed in the star-forming regions, which are seen in
the UV and optical ranges. Thus, there is no appreciable hidden SF
which is not seen in the UV range. This result is in agreement with conclusions
made by \citet{IT11,IT16} and \citet*{I09}.

\subsection{Infrared excess}

The infrared excess $IRX$ is defined as the ratio of the observed luminosity 
of dust in the IR range to the observed UV luminosity. This quantity is somewhat
arbitrary and depends on the adopted wavelength range in which the UV
luminosity is derived. For the sake of comparison with other
studies we define $IRX$ as 
\begin{equation}
IRX = \frac{L(IR)}{\nu L_\nu},
\label{IRX}
\end{equation}
where $\nu L_\nu$ are luminosities of the obscured SEDs at 
$\lambda$ = 1500\AA, which are proxies of observed luminosities.

The dependence of $IRX$ on the UV slope of the obscured SEDs
is shown in Fig. \ref{fig12}. There are no
offsets between the distributions of SFGs with high and low EW(H$\beta$)s.
The distribution of our galaxies 
is overlapped with that of a sample of local galaxies from the Galaxy and Mass
Assembly (GAMA) survey \citep[region delineated by blue dotted lines, ][]{W16} 
and SFGs from the Reference {\sl Herschel} sample 
\citep[region delineated by red dotted lines, ][]{B12}, which however lack 
objects
with steep slopes $\beta$ $\leq$ --1 and $\beta$ $\leq$ --1.5, respectively.
They also agree with the IRX values obtained by \citet{B16} and \citet{R12}
for SFGs at $z$ = 2 -- 3 (filled circle and filled triangles, respectively), and
distributions of typical galaxies at $z$ $\sim$ 1 -- 3 \citep{R10,F16}.
The distribution of our CSFGs with $\beta$ $<$ --1.5 is consistent
with dependencies obtained by \citet{M99}
(dashed line) and \citet{K04} (solid line).
On the other hand, the distribution of CSFGs with flatter slopes of beta
is located between the 
dependencies obtained by \citet{M99} and \citet{K04} on one side and the 
dependence by \citet{P98} (dash-dotted line) on the other side.

\section{Conclusions}\label{sec7}

In this paper we model apparent and absolute far-UV (FUV) and near-UV (NUV) 
magnitudes of a sample of $\sim$ 14000 compact star-forming galaxies
(CSFGs) selected from the Data Release 12 (DR12) of the Sloan Digital
Sky Survey (SDSS). These quantities were obtained using extrapolations
of spectral energy distributions (SEDs), which were obtained from fitting
the SDSS optical spectra. The modelled magnitudes are compared to observed 
{\sl Galaxy Evolution Explorer} ({\sl GALEX}) FUV and NUV
magnitudes to constrain the reddening law in the
UV range. We also derived the slopes of UV continua and efficiencies of
ionised photon production using modelled SEDs and extinction-corrected
H$\beta$ luminosities.
For the same CSFGs we derive intrinsic and absorbed luminosities 
in the UV range. These luminosities are compared to infrared (IR) luminosities,
which were obtained using {\sl Wide-field Infrared Survey Explorer} 
({\sl WISE}) fluxes at 3.4 - 22 $\mu$m
and various relations linking {\sl WISE} fluxes with fluxes at longer
wavelengths, 60 - 500 $\mu$m, obtained with the 
{\sl Infrared Astronomical Satellite} ({\sl IRAS}) and 
{\sl Herschel} observatories. Our main results are as follows.

1. We find that the differences between modelled and observed FUV and NUV
apparent magnitudes depend on the adopted reddening law, which is used to
obtain modelled apparent magnitudes from the CSFG' intrinsic SEDs in the UV 
range.
The best agreement between the apparent and modelled FUV magnitudes for CSFGs
with high equivalent widths of the H$\beta$ emission line 
EW(H$\beta$) $\geq$ 50\AA\ is found if the \citet{C89} reddening law 
with $R(V)$~=~2.7~--~3.1
is adopted while $R(V)$ = 2.7 is found for CSFGs with 
EW(H$\beta$)~$\geq$~150\AA. On the other hand, in CSFGs with 
EW(H$\beta$) $<$ 50\AA, 
the best fit is achieved for observed apparent magnitudes adopting 
the \citet{C94} reddening law or the \citet{C89} reddening law with higher 
$R(V)$ = 3.1. This 
implies that the properties of dust in SFGs with intense UV radiation, i.e. with
high EW(H$\beta$)s, are characterized by a higher fraction of small grains
resulting in a steeper reddening law. The agreement between the modelled and
observed NUV magnitudes is better for slightly higher $R(V)$ as compared to
the FUV range indicating that one-parameter reddening laws by \citet{C89}
fail to reproduce observations in the entire FUV and NUV ranges. 
We also found that observed and modelled 
FUV and NUV magnitudes are better reproduced if equal stellar
and nebular extinctions in the UV range are adopted.

2. It is found that the infrared luminosity of CSFGs in the 
observed wavelength range 8 -- 1000 $\mu$m is nearly equal to the 
luminosity absorbed at shorter wavelengths $\leq$ 1 $\mu$m, independently of
the stellar mass of the galaxy, extinction and the age of the starburst.
This indicates that our CSFGs are relatively transparent with small variations
of the extinction. There is no evident sign of considerable star formation (SF)
located in highly obscured regions, which are not seen in the UV and optical
ranges.

3. The slopes $\beta$ of the SEDs in the UV of our CSFGs
vary in the wide range from --2.5 to +1. Most of this spread ($>$ 70\% by value)
is caused by the extinction. Only $\leq$ 30\% of the spread in $\beta$ is due to
various starburst ages.

4. The log of the intrinsic efficiency of ionising photon production 
$\xi$ in our CSFGs varies in a wide range between 24.0
and 25.7 due to the bursting nature of the SF with clear offsets between
the CSFGs with high EW(H$\beta$) (young bursts) and with low EW(H$\beta$) 
(older bursts). 
Reducing to a zero burst age we eliminate the offset
between CSFGs with low and high EW(H$\beta$)s. The reduced values of 
log $\xi_0$ evenly scatter around the value of 25.8, which is close to the value
of zero age bursts. However, we argue that for studying the problem of
the reionisation of the Universe at redshifts 5 -- 10 the canonical value
of $\sim$ 25.2 globally may be used assuming that high-redshift SFGs were formed
with a constant rate over a period of several hundred Myr.

\section*{Acknowledgements}

Funding for the SDSS and SDSS-II was provided by the Alfred P. Sloan 
Foundation, the Participating Institutions, the National Science Foundation, 
the U.S. Department of Energy, the National Aeronautics and Space 
Administration, the Japanese Monbukagakusho, the Max Planck Society, and the 
Higher Education Funding Council for England. The SDSS was managed by the 
Astrophysical Research Consortium for the Participating Institutions.
Funding for SDSS-III has been provided by the Alfred P. Sloan Foundation, 
the Participating Institutions, the National Science Foundation, and the U.S. 
Department of Energy Office of Science. The SDSS-III web site is 
http://www.sdss3.org/. SDSS-III is managed by the Astrophysical Research 
Consortium for the Participating Institutions of the SDSS-III Collaboration. 
GALEX is a NASA mission  managed  by  the  Jet  Propulsion  Laboratory.
This research has made use of the NASA/IPAC Extragalactic Database (NED) which 
is operated by the Jet  Propulsion  Laboratory,  California  Institute  of  
Technology,  under  contract with the National Aeronautics and Space 
Administration.






\bsp	
\label{lastpage}
\end{document}